\documentclass{article}

\PassOptionsToPackage{numbers}{natbib}
\usepackage[main,final]{neurips_2026}

\usepackage{amsmath,amsfonts,bm}

\def\eqref#1{equation~\ref{#1}}

\def\1{\bm{1}}

\DeclareMathAlphabet{\mathsfit}{\encodingdefault}{\sfdefault}{m}{sl}
\SetMathAlphabet{\mathsfit}{bold}{\encodingdefault}{\sfdefault}{bx}{n}

\usepackage[utf8]{inputenc}
\usepackage[T1]{fontenc}
\usepackage{url}
\usepackage{graphicx}
\usepackage{booktabs}
\usepackage{multirow}
\usepackage{amsfonts}
\usepackage{nicefrac}
\usepackage{microtype}

\usepackage{xcolor}
\definecolor{violet}{HTML}{57068C}

\usepackage[colorlinks=true,
            linkcolor=violet,
            citecolor=violet,
            urlcolor=violet]{hyperref}

\colorlet{newgreen}{black}
\newif\ifshowchanges
\showchangestrue

\ifshowchanges
  \newcommand{\newcamr}[1]{{\color{newgreen}#1}}
  \newenvironment{newtext}{\color{newgreen}}{}
\else
  \newcommand{\newcamr}[1]{#1}
  
\fi

\usepackage{xcolor}
\colorlet{reddishblack}{red!50!black}

\title{Learning What's Real: Disentangling Signals from Measurement Artifacts in Multi-Sensor Data, with Applications to Astrophysics}

\author{%
  \parbox{0.95\linewidth}{\centering
    \mbox{Pablo Mercader-Perez\textsuperscript{1,*}} \quad
    \mbox{Carolina Cuesta-Lazaro\textsuperscript{2,3}} \quad
    \mbox{Daniel Muthukrishna\textsuperscript{1,4}} \quad
    \mbox{Jeroen Audenaert\textsuperscript{1}} \quad
    \mbox{V.\ Ashley Villar\textsuperscript{4,5}} \quad
    \mbox{David W.\ Hogg\textsuperscript{2,3}} \quad
    \mbox{Marc Huertas-Company\textsuperscript{6}} \quad
    \mbox{William T.\ Freeman\textsuperscript{1}}
    \\[6pt]
    {\normalfont
    \mbox{\textsuperscript{1}Massachusetts Institute of Technology} \quad
    \mbox{\textsuperscript{2}New York University} \quad
    \mbox{\textsuperscript{3}Flatiron Institute, Simons Foundation} \quad
    \mbox{\textsuperscript{4}Center for Astrophysics $\mid$ Harvard \& Smithsonian} \quad
    \mbox{\textsuperscript{5}NSF AI Institute for Artificial Intelligence and Fundamental Interactions (IAIFI)} \quad
    \mbox{\textsuperscript{6}Instituto de Astrofísica de Canarias}}
  }%
}

\begin{document}

\maketitle

\begingroup
\renewcommand\thefootnote{*}
\footnotetext{Correspondence: \texttt{pablomer@mit.edu}. Code: \url{https://github.com/pablomerc/galaxy-counter}.}
\endgroup

\begin{abstract}
Data collected from the physical world is always a combination of multiple sources: an \newcamr{intrinsic} signal from the physical process of interest and a signal from measurement-dependent artifacts \newcamr{introduced by} the sensor or instrument. This secondary signal acts as a confounding factor, limiting our ability to extract information about the underlying physics. Moreover, it poses significant challenges for combining data in heterogeneous or multi-instrument \newcamr{settings}. To disentangle these factors of variation, we propose a dual-encoder architecture with a counterfactual generation objective that leverages overlapping observations. The resulting representations explicitly separate intrinsic signals from sensor-specific distortions and noise, and can be used for counterfactual view generation, parameter inference, and instrument-independent similarity search---all unconfounded by measurement artifacts. We demonstrate the effectiveness of our approach in a multi-instrument setting on astrophysical galaxy images from the DESI Legacy Imaging Survey and the Hyper Suprime-Cam Survey. This framework provides a general recipe for scientific self-supervised pretraining: construct training pairs from overlapping observations of the same physical system, treat sensor- or modality-specific effects as augmentations, and learn invariant representations through counterfactual generation.

\end{abstract}
\section{Introduction}
\label{intro}

Observational data collected from the physical world can be viewed as the result of a causal process. First, an event of interest occurs, generating a physical signal. For example, a distant pulsating star emits electromagnetic waves, a person speaking creates pressure waves in the air, and a beating heart produces pressure waves that propagate through blood vessels. %
This signal cannot be directly observed. Instead, it must pass through a sensor or instrument that records it. Any instrument used introduces measurement artifacts, often referred to as \textit{instrument systematics} or bias. These effects include transformations like the nonlinearity of Charge-Coupled Devices (CCDs) in a camera or a microphone's frequency response. %
These act as \textit{confounding} factors. However, with sufficient understanding of the measurement system, they can often be partially modeled and disentangled from the underlying signal. 
Additionally, stochastic noise imposes a limit on measurement precision.

In other words, \textit{$\text{Observation} = f(\text{Signal},\text{Instrument}) + \text{Noise}$}. In some cases, $f$ is relatively simple, such as a point-spread function (PSF) convolution that only requires accurate characterization of the optics.
However, it often cannot be modeled fully analytically.
The true signal is explained by variables that are independent of the measurement and contain the scientifically relevant information; we refer to these as \textbf{physics variables}. The instrumental effects arise from measurement distortions and act as confounders; we refer to these as \textbf{instrument variables}.

We present a general approach to disentangle physics and instrument variables when the same system is observed by multiple instruments. The key insight is that cross-matched observations form natural pairs of \emph{same physics, different instrument}, providing a training signal for separating the two sets of factors without requiring an explicit forward model of either instrument.

We demonstrate our approach on astrophysical galaxy images, a domain where modeling instrument systematics is critical. Photons emitted by a distant source must traverse cosmic dust and the Earth's atmosphere before being recorded by ground-based telescopes, each with its own optics, detector characteristics, and observing conditions. As a result, the same galaxy can look remarkably different across two telescopes due to differences in resolution, noise, and calibration. Historically, obtaining observations of the same phenomena with multiple instruments has been essential for distinguishing real physical signals from instrumental artifacts. 
In this work, we train our model on ${\sim}100{,}000$ cross-matched galaxy images from two different ground-based telescopes and show that we learn disentangled representations separating the intrinsic properties of galaxies from the instrument-specific effects, enabling robust inference of galaxy properties, outlier detection, counterfactual generation across instruments, and instrument-independent similarity search.

\begin{figure}[t]
    \centering \includegraphics[width=1\linewidth]{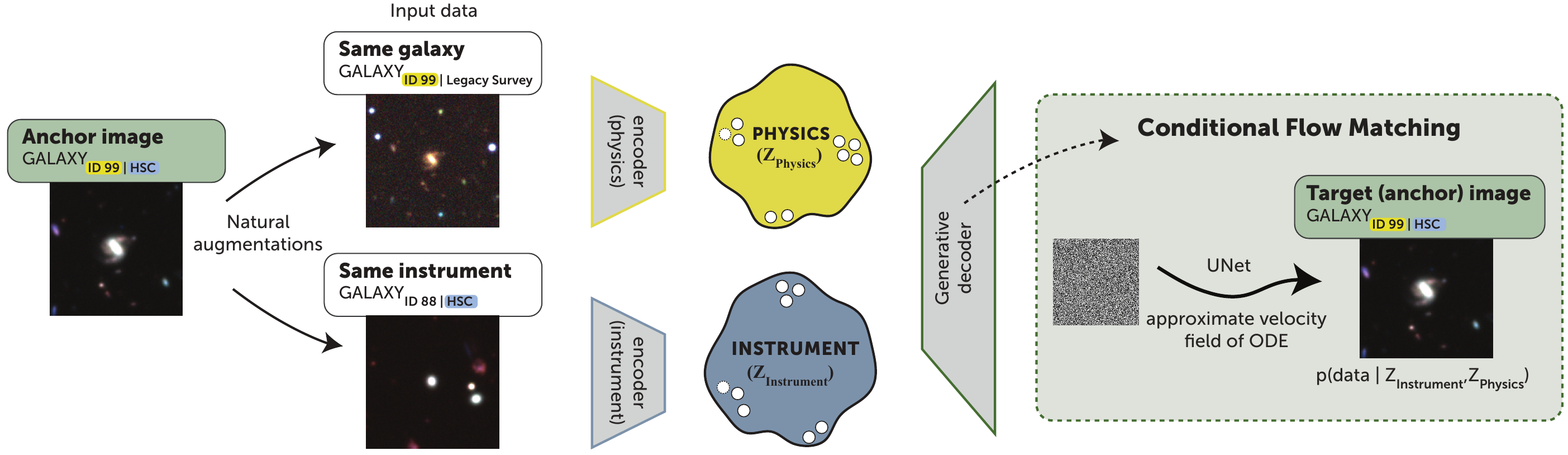}
    \caption{\textbf{Counterfactual Reconstruction.} The model learns to disentangle intrinsic galaxy properties from instrument systematics by reconstructing an anchor image. Training uses \textbf{data triplets}: an \textbf{anchor observation} (signal source $s$ from instrument $i$), an instrument-augmented observation (\textbf{same source} $s$, a different instrument $i'$), and a physics-augmented observation (a different source $s'$, \textbf{same instrument} $i$). The \textit{physics} encoder extracts latents from the instrument-augmented pair to capture invariant physical features, while the \textit{instrument} encoder processes the physics-augmented observation \newcamr{to capture instrument-specific effects}. These latent variables jointly condition the flow-matching decoder. Crucially, the anchor image is never fed into the encoders; it is used only as a target for the flow-matching loss, forcing the model to generate the reconstruction counterfactually.}
    \label{fig:architecture}
\end{figure}

\section{Related work}
\textbf{Disentangled representation learning.} Learning representations that separate independent factors of variation is a longstanding goal in representation learning. \citet{Tenenbaum99} explored bilinear representations to separate style and content. Early work on variational autoencoders (VAEs), such as $\beta$-VAE \citep{higgins2017betavae}, encouraged disentanglement by regularizing the latent space via the loss function. %
A natural source of structure is multi-view observations, where the same underlying content is observed under varying conditions or at different times. %
From a causal perspective, \citet{Scholkopf2021} framed disentanglement as the recovery of independent causal mechanisms, connecting representation learning and causal inference. Our framework draws on these ideas but enforces structural disentanglement through a specific architecture-driven information bottleneck and a counterfactual generation objective, rather than relying on hand-engineered loss functions. %

\textbf{Foundation models in the sciences.} Pre-training self-supervised representations on large scientific datasets has recently seen success in different areas of science like biology \citep[e.g.,][]{Zeming2023, ross2022large} and satellite imagery \citep[e.g.,][]{cong2022satmae}. In astrophysics, \citet{walmsley2022galaxyfoundationmodelshybrid} demonstrated the value of self-supervised representations for downstream tasks. More recently, \citet{parker2025aion1omnimodalfoundationmodel} developed AION-1, a large-scale foundation model trained via masked modeling across modalities such as imaging and spectra from the Multimodal Universe dataset \citep{themultimodaluniversecollaboration2024multimodaluniverseenablinglargescale}. However, a limitation common to these approaches is that instrument-specific effects are either ignored or treated as independent modalities without explicit disentanglement. This has practical consequences for downstream tasks. For instance, a model developed for ESA's Euclid mission \citep{EuclidSiudek2025} found that instrument systematics, rather than physically interesting sources, dominated the outliers identified in the latent space. Our work addresses this by building the separation of physics and instrument factors into the architecture itself.

\textbf{Incorporating multi-instrument observations in foundation models.} Recently in astrophysics, \citet{audenaert2025causalfoundationmodelsdisentangling} leveraged the causal representation ideas from  \citet{Scholkopf2016,Scholkopf2021,Hattori2022}, to causally separate astrophysical observations into a physics and instrument component by using a dual-encoder architecture. They used a dataset of simulated astronomical time series captured by multiple instruments to learn a \textit{physics latent space} that encodes information about the target star and an \textit{instrument latent space} that encodes information about the measurement configuration. This is achieved by leveraging triplets of overlapping observations and using contrastive learning. %
While effective, contrastive objectives have inherent limitations for our setting. By aligning representations across views of the same source, they tend to retain only the information shared across all views, effectively collapsing toward the lowest-resolution observation in the set and discarding fine-grained features that are visible in higher-quality instruments. Also importantly, contrastive embeddings are not optimized for pixel-level reconstruction and producing counterfactual views would require training a separate generative model on the learned embeddings, with no guarantee that the necessary information is preserved. Our generative objective addresses these limitations.

\section{Methods}
We present a counterfactual generation objective designed to learn representations that disentangle physical properties from measurement systematics. Our goal is to capture robust physical information into a physics latent space, while learning a data-driven noise model for each of the instruments.

\textbf{Counterfactual generative objective.} \label{sec:counter}
Our framework leverages large-scale scientific datasets with overlapping observations from multiple instruments. In this setting, natural augmentations can be used to construct triplets: an anchor, an observation of the same source (but different instrument), and an observation on the same instrument (but different source), as shown in Fig.~\ref{fig:architecture}. %

A \textit{physics encoder} receives a sample from the target source but from a different instrument, so it is encouraged to reduce the loss by capturing the underlying physical properties of the source and becoming invariant to the instrumental distortions. Similarly, an \textit{instrument encoder} receives a sample from the target instrument but a different source, so it is encouraged to capture instrument-specific distortions while ignoring the content specific to the source. The decoder combines these two latent embeddings to reconstruct an unseen target or anchor observation.

Formally, consider a set of $N$ sources observed across $M$ instruments, where $x_{j,k}$ denotes the observation of source $j$ as captured by instrument $k$. We aim to learn the conditional distribution
$p\!\big(
  x_{j,k} \,\big|\,
  \{ z_{\text{phy}}(x_{j,k'}) \}_{k' \neq k},\,
  \{ z_{\text{ins}}(x_{j',k}) \}_{j' \neq j}
\big),
$
for all source--instrument pairs $(j,k)$. Here $ \{ z_{\text{phy}}(x_{j,k'}) \}_{k' \neq k}$ are the physics embeddings extracted from observations of the \emph{same source} $j$ by \emph{other instruments} $k' \neq k$, which provide information about the source's intrinsic physical properties. The set $\{z_{\text{ins}}(x_{j',k})\}_{j' \neq j}$ are the instrument embeddings extracted from observations of \emph{other sources} $j'$ by the \emph{same instrument} $k$. This formulation allows the model to leverage all available multi-instrument data: the physics of the target source is informed by its appearances across different instruments, while the instrument response is informed by how other sources appear under similar observing conditions.
\newcamr{We concatenate the physics and instrument embedding tokens into a variable-length sequence for decoder cross-attention, allowing the number of source and instrument neighbors to vary (see the ``Model implementation'' paragraph in Section~\ref{sec:applicationtoastro} and Appendix~\ref{app:architecture}).}

We model this conditional distribution using flow matching \citep{lipman2023flowmatchinggenerativemodeling}, learning a velocity field $u_\theta(x, t, z_{\text{phy}}, z_{\text{ins}})$ that transports samples from Gaussian base distribution to the target distribution. The training loss is given by the error on the predicted velocity field,
\begin{equation}
    \small
    \mathcal{L}_{\text{FM}}(\theta) = \mathbb{E}_{\substack{t \sim \mathcal{U}[0,1] \\ j \sim [N], k \sim [M] \\ x_{j,k} \sim p_{\text{data}} \\ x_0 \sim \mathcal{N}(0, I) }} \left[ \left\| u_\theta\left(x_t, t, \{z_{\text{phy}}(x_{j,k'})\}_{k' \neq k}, \{z_{\text{ins}}(x_{j',k})\}_{j' \neq j}\right) -  (x_{j,k} - x_0) \right\|^2 \right]
\end{equation}
where $x_t = (1-t)x_0 + t x_{j,k}$. That is, we sample a source $j$, a target instrument $k$, a noise sample $x_0$, and a timestep $t$, and train the velocity field to reconstruct the observation $x_{j,k}$ from a noisy state $x_t$ conditioned on physics latents from other instruments and instrument latents from different sources. 

Once trained, the physics embedding supports downstream tasks such as outlier detection, parameter inference, classification, or embedding-based retrieval, while the instrument space allows to find similar systematics and observing conditions. The generative decoder enables generating counterfactual views. In astrophysics, a particularly promising use case is to inform follow-up observations by predicting high-resolution observations from expensive facilities using low resolution observations from a different instrument.

\section{Application to Astrophysics: Galaxy foundation models}\label{sec:applicationtoastro}

Datasets in astrophysics have grown considerably over the last decade. In particular, galaxy surveys such as DESI, have now imaged over a billion galaxies while taking spectra of over 10 million galaxies. This has sparked interest in developing foundation models for galaxies that learn semantic representations of astronomical data, helping us search large databases~\citep{Koblischke_2026}, identify interesting anomalies, and predict galaxy properties of interest. Up to now, these models have treated data from different instruments as independent modalities \citep{parker2025aion1omnimodalfoundationmodel}. Here, we show that counterfactual reconstruction improves robustness to instrument systematics while enabling new foundational capabilities, including cross-instrument generation, instrument-independent retrieval, and data-driven denoising.

We take a cross-matched dataset from the Multimodal Universe
that includes $\sim$100,000 cross-matched galaxy images from the DESI Legacy Imaging Surveys \citep[Legacy,][]{Dey_2019} and the Hyper Suprime-Cam \citep[HSC,][]{Aihara_2017} survey as a representative multi-instrument setting. HSC is a deeper and higher-resolution instrument that can detect significantly fainter features that would otherwise be masked by noise in Legacy. However, by making this compromise, Legacy covers a larger area of the sky, $\approx 20{,}000 \, \text{deg}^2$, compared to HSC, $\approx 1{,}200 \, \text{deg}^2$, resulting in a trade-off between coverage \newcamr{and image quality
(both photometric depth and angular resolution).}

\textbf{Data preprocessing.} We have two sets of heterogeneous flux measurements: HSC provides five color filters with discrete wavelength bands ($g, r, i, z, y$), and the Legacy Survey uses four ($g, r, i, z$). Despite the shared designations, they differ significantly in their central wavelengths, pixel scales, zero-points, and noise characteristics. Our preprocessing pipeline follows the strategy established by \cite{parker2025aion1omnimodalfoundationmodel}, with one key difference: rather than treating all nine bands as distinct channels, we align the overlapping $(g, r, i, z)$ filters across both surveys into a unified four-channel representation and treat both sources as the same modality. For this purpose, we discard the additional HSC $y$-band channel, but future work could incorporate it.%

Following \cite{parker2025aion1omnimodalfoundationmodel}, we normalize the zero-points by rescaling HSC measurements to the Legacy standard of 22.5 mag using the relation: $s = 10^{(ZP-22.5)/2.5}$. Then, we apply an $\text{arcsinh}$ normalization to handle the images' large dynamic range. Finally, we crop the HSC images into $48 \times 48$ pixel cutouts. To account for the differing pixel scales between surveys ($0.168^{\prime\prime}$ for HSC vs.\ $0.262^{\prime\prime}$ for Legacy), we extract the inner $31 \times 31$ pixels of the Legacy images and upsample them to $48 \times 48$ pixels via linear interpolation; this ensures spatial alignment and a consistent field of view across pairs.

\textbf{Data triplets.} In this dual-survey setting, each data point has a unique galaxy pair, providing the cross-survey view supplied to the physics encoder. For the instrument views, we condition on up to five images of different galaxies observed by the same survey as the anchor. We select these as the five nearest spatial neighbors in the given survey within an angular separation of $3$ arcminutes. This choice is motivated by the fact that instrumental systematics - such as PSF size and depth - vary smoothly across the sky, so spatially nearby observations share similar instrument conditions with the anchor. 

\textbf{Model implementation.} We use a modified ResNet-18 \citep{he2015deepresiduallearningimage} for both encoders (physics and instrument) as described in Appendix~\ref{app:architecture}. We map each input image to a sequence of 4 spatial tokens of 16 dimensions, i.e. a latent representation of shape $(B, 4, 16)$. To accommodate a variable number of conditioning examples, we treat these embeddings as a set of tokens. For a given input set, the tokens from all images are concatenated into a single sequence of variable length. Then, we use cross-attention layers within our velocity network, a \texttt{UNet2DConditionModel} \citep{von-platen-etal-2022-diffusers}. This allows the model to leverage an arbitrary number of context image embeddings during the generative process.

\subsection{Disentangled representations}
\begin{figure}[t]
    \centering
    \includegraphics[width=0.94\linewidth]{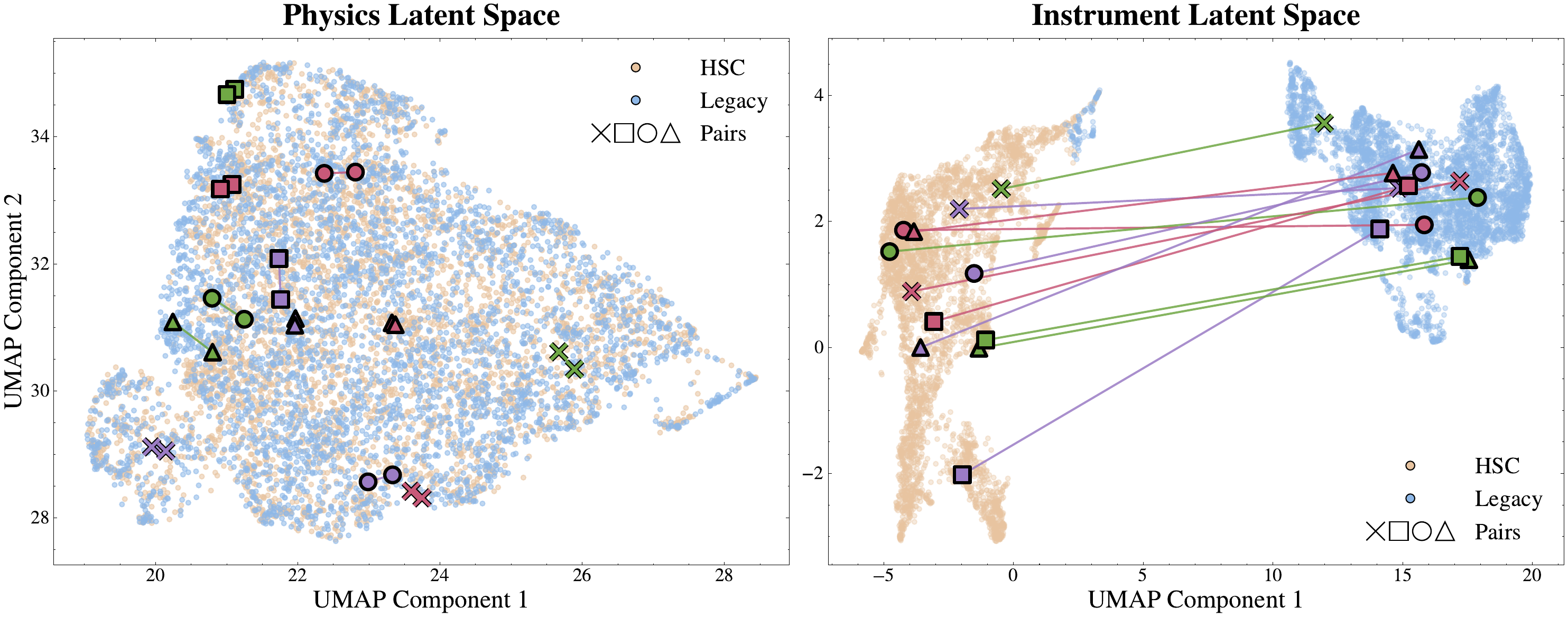}
    \vspace{-5pt}
    \caption{\textbf{Disentanglement.} UMAP projections of the physics (left) and instrument (right) latent spaces. Orange and blue points represent HSC and Legacy images, respectively. Matched markers ($\triangle$, $\times$, $\square$, $\circ$) denote cross-survey pairs.
    In the \textit{physics space}, the encoders produce overlapping distributions where cross-survey pairs are mapped to similar coordinates. In contrast, the \textit{instrument space} shows a clean separation between surveys and pairs, confirming that this space captures survey-specific features rather than intrinsic galaxy properties. \newcamr{A complementary quantitative evaluation of downstream $R^2$ scores for physical and instrumental properties is shown in Fig.~\ref{fig:r2-scores}.}}
    \vspace{-10pt}
    \label{fig:galaxy-latents-umap}
\end{figure}
First, we demonstrate that the physics and instrument latent spaces do show the expected disentanglement. We encode $8{\small,}192$ image pairs from HSC and Legacy Surveys using both the physics and instrument encoders, and visualize the resulting latent embeddings using a 2D UMAP projection \citep{mcinnes2020umapuniformmanifoldapproximation} in Fig.~\ref{fig:galaxy-latents-umap}. The latent space demonstrates a clear separation of information. In the instrument space, HSC and Legacy observations form two distinct clusters, confirming that the encoder successfully captures survey-specific characteristics. In the physics space, the two surveys have largely overlapping distributions, indicating that the encoder has learned an approximate domain-invariant representation of the underlying galaxy properties. The overlap is, however, not expected to be perfect since the two surveys provide different amounts of constraining power on the underlying physical properties. %
The emergence of alignment of pairs observed in Fig.~\ref{fig:galaxy-latents-umap} from the reconstruction objective alone confirms that the architectural bottleneck is sufficient to drive the convergence of physics representations across the two surveys.

\textbf{Outlier detection unconfounded by instrument systematics.} Naively learned representations often bias outlier detection toward rare instrument artifacts rather than physically interesting sources. Our disentangled representation addresses this by separating physics and instrument factors, enabling outlier detection separately in each latent space: rare astrophysical sources in the physics space and rare observing conditions in the instrument space. %

We encode all $\sim$100{,}000 HSC galaxies using the physics encoder and the instrument encoder separately. As a baseline, we encode the same images with AION-1 \citep{parker2025aion1omnimodalfoundationmodel}, a foundation model that treats each survey as an independent modality without explicit disentanglement. For each of the three latent spaces, we fit a normalizing flow to the empirical distribution of embeddings and rank all sources by their estimated log-likelihood (Appendix~\ref{app:flow}). The lowest-likelihood examples are the candidate outliers.

\begin{figure}[t]
    \centering
    \includegraphics[width=0.99\linewidth]{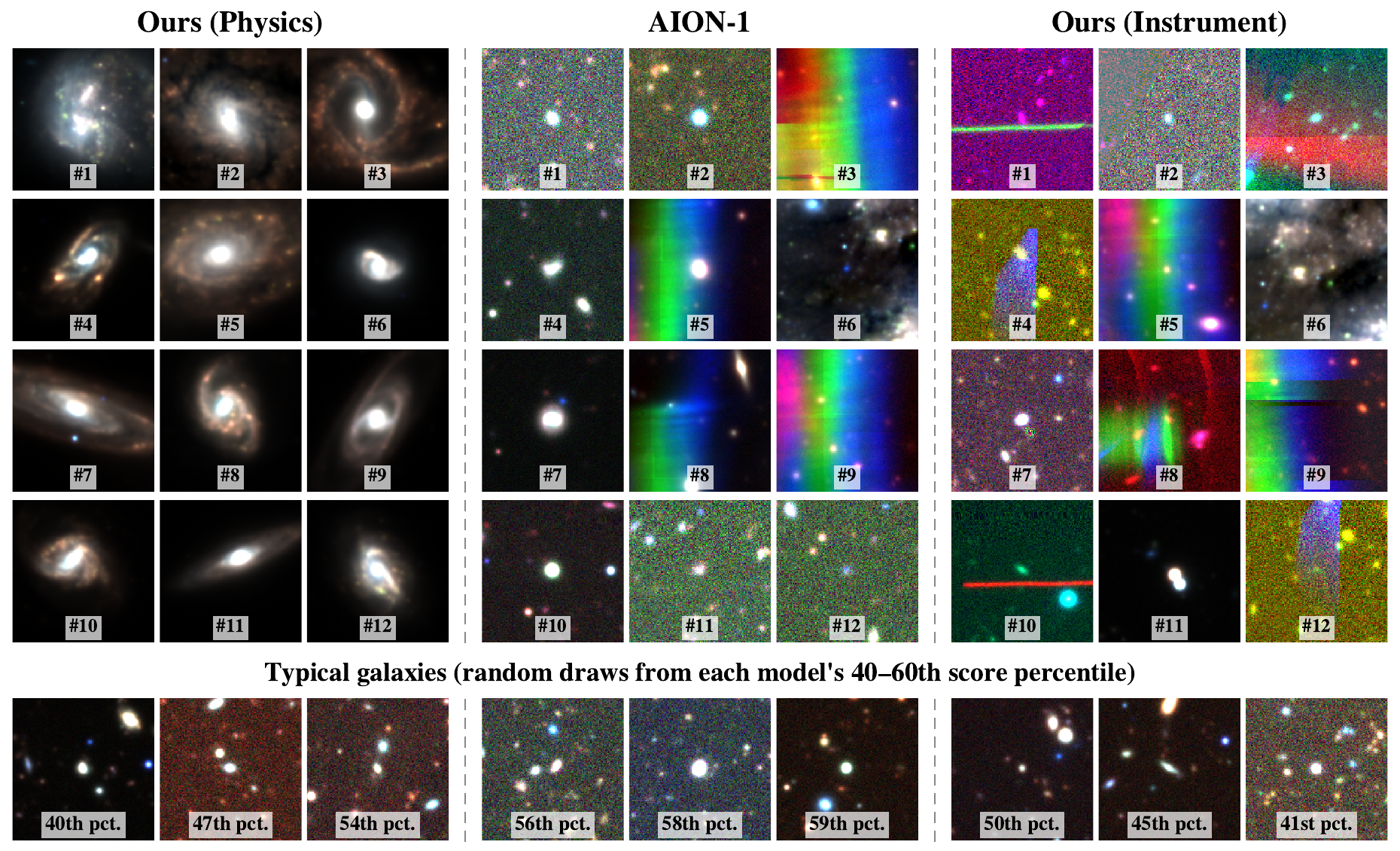}
    \caption{\textbf{Outlier Detection.} Top twelve outliers from each of the three latent spaces. \textbf{Left:} our physics latent reveals morphologically unusual sources (e.g., galaxy mergers), free of artifacts. \textbf{Middle \& Right:} anomalies from AION-1 and our instrument latent are dominated by instrumental pathologies (saturated pixels, bands, noise patterns), confirming that our representation separates physical anomalies from instrument systematics.
    \newcamr{\textbf{Bottom row:} for reference, three galaxies drawn uniformly at random from the 40--60th percentile of each model's anomaly score. Typical galaxies look alike in all three spaces.}}
    \vspace{-12pt}
    \label{fig:anomalies}
\end{figure}

Fig.~\ref{fig:anomalies} shows the top twelve outliers from each latent space. The AION-1 anomalies contain several instrumental pathologies: saturated pixels from bright stars, vertical rainbow bands, and high-noise patterns. Outliers from our instrument space show similar characteristics, confirming that the instrument encoder captures systematics-related variation. In contrast, anomalies from our physics space are visibly free of these artifacts and reveal morphologically unusual sources: galaxy mergers, disturbed systems, and other candidate rare objects. \newcamr{Notably, galaxies sampled near the median anomaly score appear qualitatively similar across the three latent spaces, while their highest-ranked outliers are visibly different.}
We show cutouts at the full $160$$\times$$160$ pixel size. 

\begin{figure}
    \centering
    \includegraphics[width=0.9\linewidth]{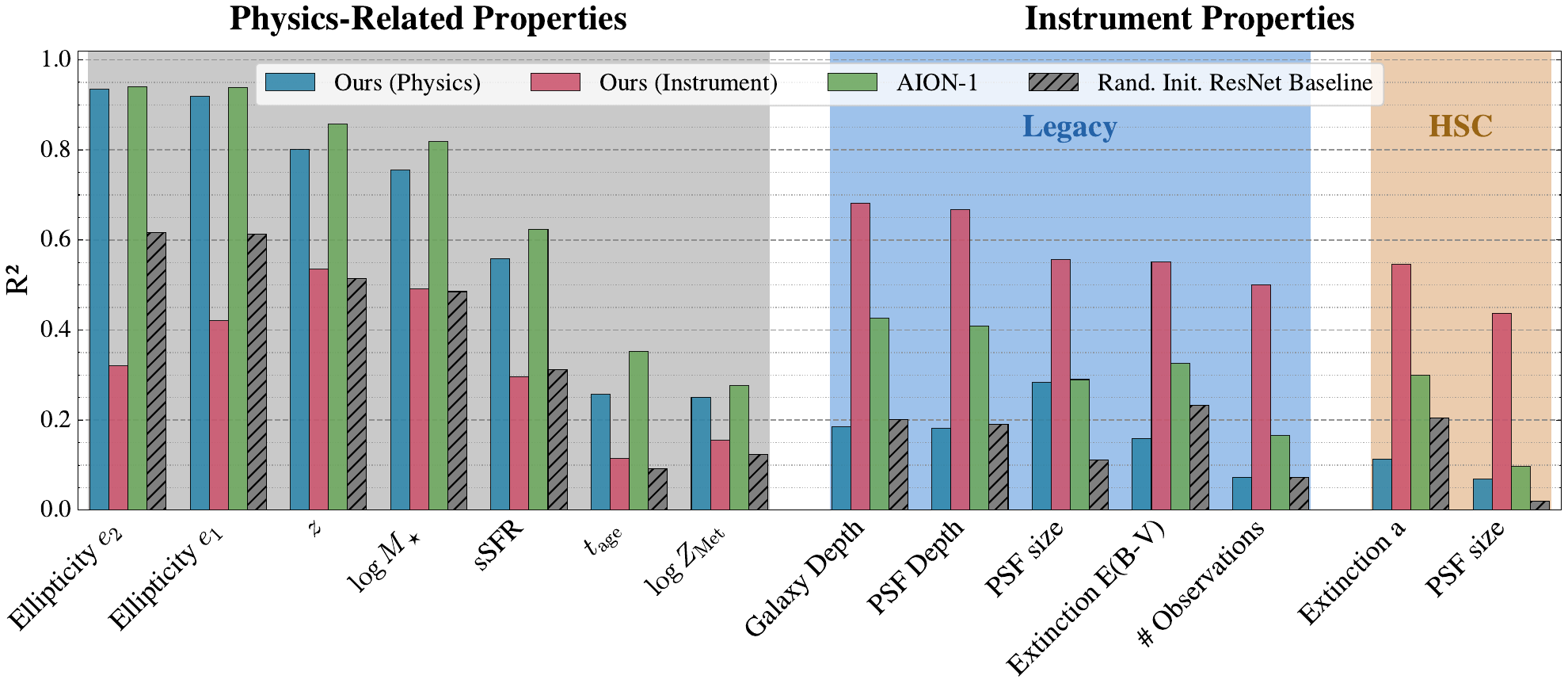}
    \vspace{-6pt}
    \caption{ \textbf{Probing Latent Disentanglement via Downstream Regression.} We report $R^2$ scores for the prediction of physics-related properties (left) and instrumental properties (right) on four sets of representations: our physics embeddings (blue), our instrument embeddings (red), AION-1 embeddings which treat each survey as an independent modality (green), and a randomly initialized ResNet-18 of the same architecture as our encoders (hatched black). We distinguish physics-related properties and instrument properties, which are grouped by survey. The asymmetry between the two panels demonstrates that our architecture successfully separates physics properties from instrument-specific information.}
    \vspace{-6pt}
    \label{fig:r2-scores}
\end{figure}

\textbf{Parameter inference robust to instrument systematics.} 
To evaluate the information content and disentanglement of our latent spaces, we train a small MLP (Appendix~\ref{app:MLP}) to predict physical and instrumental properties from embeddings extracted by each encoder on Legacy and HSC images. For the physics latent space we flatten across the token dimension; for the instrument latent space we apply mean pooling. In both cases, we concatenate the HSC and Legacy embeddings before feeding them into the MLP. We compare against AION-1 (Base) embeddings and against a frozen, randomly initialized ResNet-18 with the same architecture as our encoders. Fig.~\ref{fig:r2-scores} shows the resulting regression $R^2$ per property.

The physical properties of galaxies were inferred by \citet{Hahn2023} from spectroscopic data, not from the images themselves, so they are only partially recoverable from imaging alone.
Moreover, we note that these are not ``physics only" properties. For instance, the galaxy's redshift determines its apparent brightness, angular size, and observed colors---properties that are also changed by instrument conditions such as depth and PSF. Morphology is also affected by the instrument's PSF. Therefore, the instrument encoder will inevitably retain some of this information. On the other hand, observing conditions, such as PSF size and depth, vary across each survey's footprint as a function of sky position, determined by the tiling strategy and the atmospheric conditions during each field's observation. Therefore, the physics latent space may retain some of this information. Details on the properties and source catalogs are provided in Appendix~\ref{appendix:gal}.

The physics latent space recovers physical properties at a level comparable to AION-1. The randomly initialized ResNet serves as a reference for what an encoder of this architecture captures by default, purely from the inductive biases of the ResNet \citep{10.5555/3104482.3104619}, without any training signal encouraging or discouraging it to encode specific information. Comparing our trained encoders to this baseline isolates what each latent space has actually been pushed to learn. Under this lens, the instrument latent space retains some information about physical properties, in particular redshift, stellar mass, and morphology. As explained above, this is expected to some degree, though we cannot quantify it. In the future, we will study how the capacity of the instrument bottleneck affects these results, especially given that the untrained ResNet exceeds our encoder's predictive power on some properties.

To further disentangle genuine leakage of instrument into physics from shared spatial structure, we train a ResNet to predict each survey's instrument properties directly from the other survey's image ("Cross-Predict ResNet"). This cross-prediction baseline measures the shared spatial component: any predictive power it achieves reflects information that is inherently common to both surveys and therefore accessible to either encoder by construction. We find that predictions from physics latents are always worse than the cross-predict baseline, verifying that the physics latent space is not capturing private instrument information. The full results are shown in Appendix~\ref{app:downstream}.

The key diagnostic is the asymmetry in the split of properties: the physics latents substantially outperform instrument latents at predicting (mostly) physical properties, whereas the pattern reverses for instrument properties. 
Notably, our latent spaces often contain less information than the baselines when used for cross-prediction (e.g. predicting physics-related properties from the instrument latents), indicating that the model has actively learned to erase instrument-specific information from the physics space and vice versa. On the other hand, we note that AION-1 embeddings retain sensitivity to instrumental properties, but our instrument latent space outperforms both baselines in predicting instrument-related properties. This highlights the primary advantage of our architecture: by explicitly disentangling the latent factors, we achieve a more efficient separation of physics and instrumental properties than monolithic foundation models. 

\begin{figure}
    \centering
    \includegraphics[width=1\linewidth]{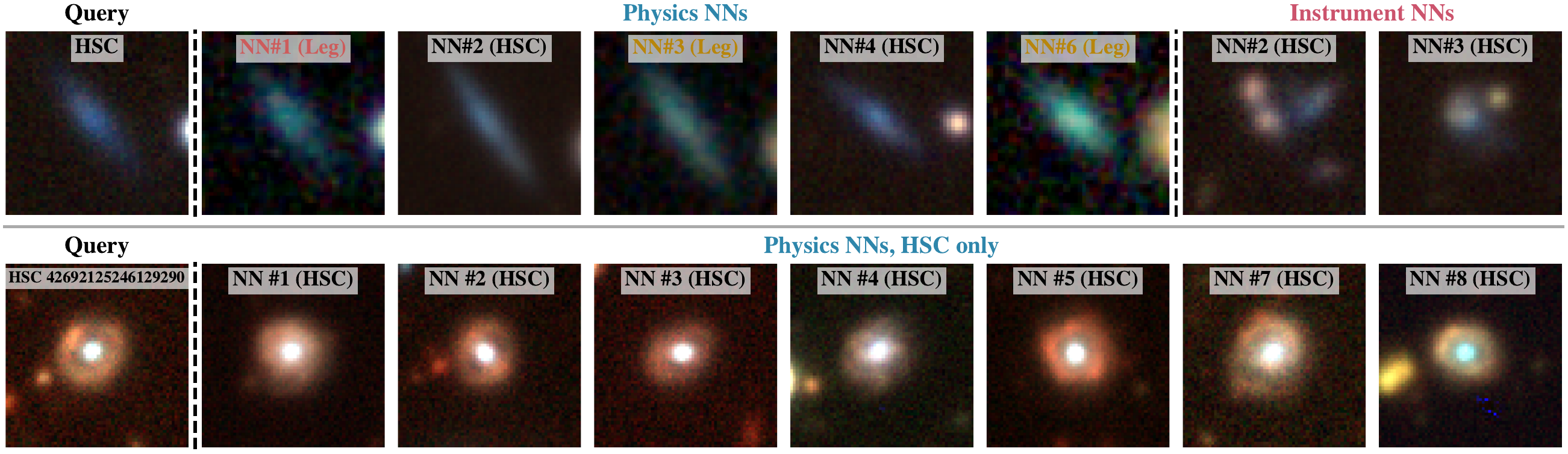}
    \caption{\textbf{Instrument-Invariant Search and Rare-Object Discovery.} \textbf{Top (Cross-Survey Retrieval):} We search the latent spaces using an HSC observation as query. \textit{Physics Space:} The pair from the alternate survey is identified as the top-1 neighbor, with additional top-ranked neighbors showing physically similar galaxies from a mixture of both instruments. Notably, the four examples shown could have been found within the top nine nearest neighbors independent of which survey the query came from. \textit{Instrument Space:} Neighbors in the instrument space show similar noise characteristics independent of the underlying galactic signal. \textbf{Bottom (Strong Lenses):} Nearest neighbor search within the physics latent space using a known strong lens as query, focusing only on the HSC images. The retrieved top neighbors show features suggestive of strong lensing, demonstrating the model's capacity for rare-object discovery.}
    \vspace{-10pt}
    \label{fig:nn}
\end{figure}

\textbf{Instrument-independent similarity search and rare-object discovery.} We can also use the learned embeddings from our model to search for nearest neighbors (NNs) in the physics and instrument spaces. In astrophysics, \newcamr{we often search large datasets for rare objects}, such as galaxy mergers or strong lenses.
By performing an $L_2$ nearest-neighbor search over the encodings generated by our model for our dataset of ${\sim}100$k cross-matched galaxies, we demonstrate that querying the physics space retrieves physically similar objects from either survey, whereas querying the instrument space isolates objects with similar noise properties. Fig.~\ref{fig:nn} \textbf{(Top)} shows \newcamr{an example where an HSC image is used as the query}. A mixture of HSC and Legacy images is found in the top neighbors in the physics space. In fact, the four examples shown could have been found among the top nine NNs, even if we used the Legacy image of that galaxy instead. This demonstrates the ability to perform \textit{instrument-independent similarity search} in physics space.

To further demonstrate the effectiveness of our disentangled physics representation, we also query the physics latent space with objects from a rare class. Fig.~\ref{fig:nn} \textbf{(Bottom)} shows an example where the query is a known strong gravitational lens. The top eight nearest neighbors successfully retrieve multiple HSC images that exhibit strong morphological evidence to be considered candidate lenses. This confirms that the physics latent space effectively isolates and groups complex physical features, streamlining the discovery of rare astrophysical phenomena.

\subsection{Counterfactual cross-instrument generation.}

\begin{figure}
\centering \includegraphics[width=0.9\linewidth]{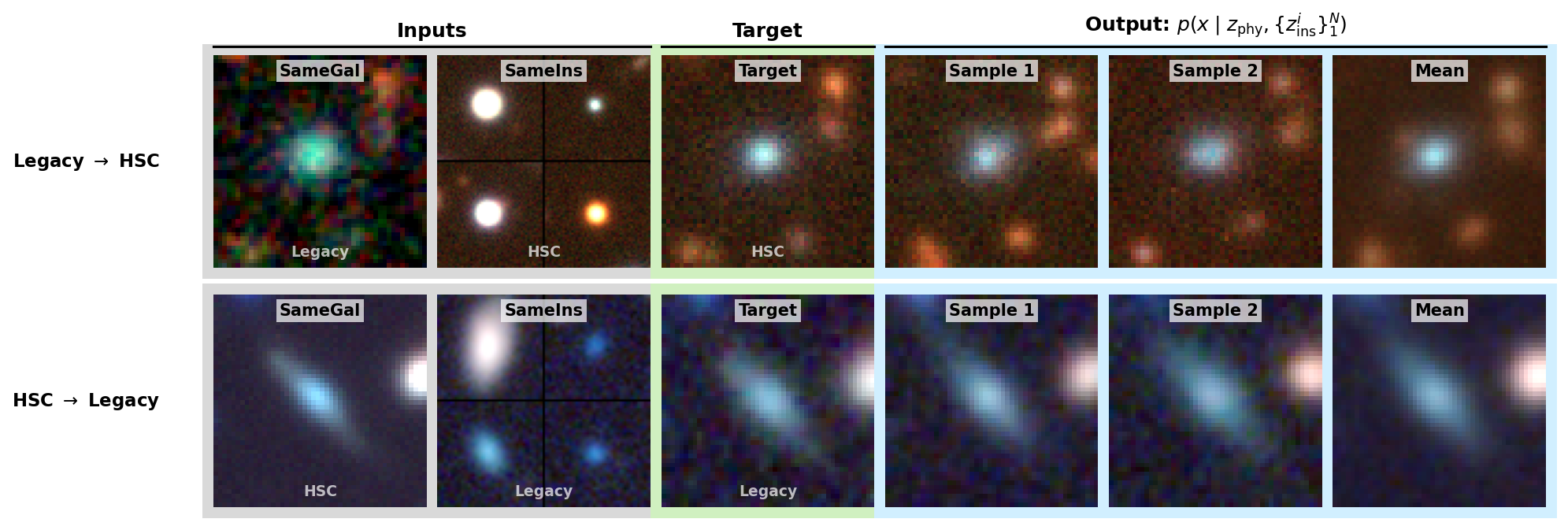} 
\caption{\textbf{Cross-instrument Galaxy Reconstructions.} \textbf{Columns 1--2 (Inputs):} The target galaxy observed via an alternate instrument (input to the physics encoder) and a set of up to five different galaxies imaged by the target instrument (input to the instrument encoder). \textbf{Column 3 (Ground Truth):} The original target image. \textbf{Columns 4--5 (Posterior Samples):} Independent samples generated by the decoder, demonstrating the model's ability to capture noise and uncertainty. \textbf{Column 6 (Posterior Mean):} The pixel-wise mean of the generated posterior samples.} 
\vspace{-10pt}
\label{fig:galaxy-reconstructions} \end{figure}

 Fig.~\ref{fig:galaxy-reconstructions} shows the model's ability to reconstruct target observations for both surveys and produce diverse posterior samples that capture the inherent uncertainty in the mapping. In addition to being our training objective, we can use counterfactual generation to prioritize follow-up observations. The discovery of rare objects, such as strong gravitational lenses, often depends critically on image quality. HSC's higher-resolution allows to resolve features, such as faint arcs, that are blurred or undetected in Legacy imaging. Our model can generate the expected HSC-like appearance for any galaxy in the Legacy footprint, effectively acting as a learned survey simulator that performs instrument-aware super-resolution and denoising. This could enable more targeted follow-up: one could first generate counterfactuals views to identify the most promising objects, those where the model predicts lensing features that are plausibly present but partly unresolved in the Legacy data. The actual follow-up observation then provides the real photons needed for confirmation.

More broadly, this approach could support population-level studies by extending HSC-like morphological analysis across the full $\sim$20,000 deg² Legacy footprint. However, we caution that the model generates the statistically expected appearance given the Legacy input, not a true observation. It can sharpen and denoise features that are marginally detected, but it cannot recover information that is entirely absent from the input. Features well below the Legacy detection threshold will not appear in the counterfactual view. The generated images are best understood as informed predictions that reduce the search space for follow-up, rather than replacements for real, deeper observations.
To evaluate the generated images, we report MSE computed on the generated images. The posterior samples achieve an MSE of $0.081$ for HSC-anchored and $0.197$ for Legacy-anchored reconstructions, evaluated over $256$ held-out galaxies.

\textbf{Zero-shot pipeline transfer.}
To further validate that our counterfactual HSC images preserve intrinsic galaxy morphology, we train a ResNet-18 model to predict galaxy ellipticities using 25,000 real HSC images, and apply this frozen model to counterfactual HSC images generated from Legacy observations of held-out galaxies. We found that the predictions match the ground truth with near-identical accuracy for both real ($R^2 = 0.82$) and generated ($R^2 = 0.81$) images. This demonstrates that counterfactually generated images can serve as drop-in replacements for real data, enabling the direct reuse of existing inference pipelines across surveys (Fig.~\ref{fig:generative_metrics}, right; Appendix~\ref{app:generative-eval}). %

\begin{figure}[htbp]
    \centering
    \includegraphics[width=\linewidth]{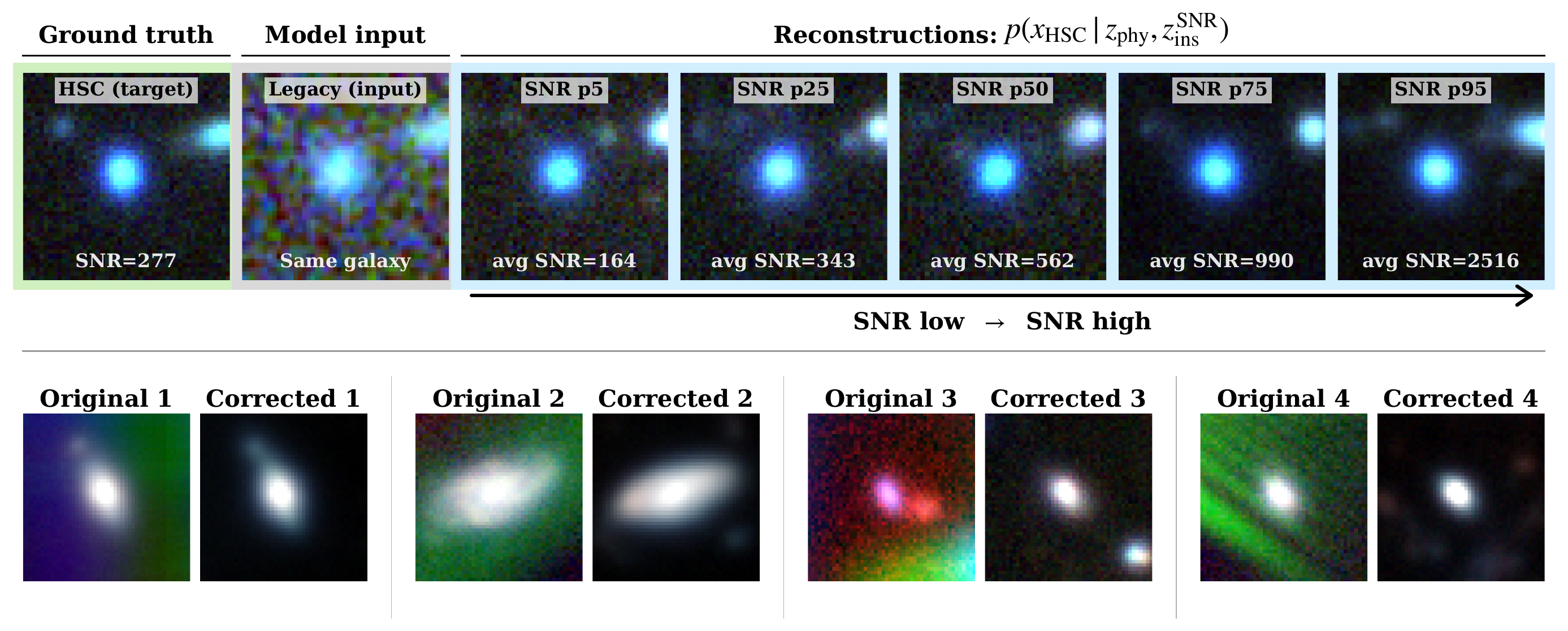}
    \vspace{-8pt}
    \caption{\textbf{Data-driven Noise Model.} \textbf{Top (Progressive Denoising):} Reconstructions of a target galaxy from its Legacy counterpart, varying the instrument conditioning across different percentiles of the HSC SNR distribution. The model successfully generates progressively cleaner images; the average SNR shown is that of the images used to condition the reconstruction. \textbf{Bottom (Systematic Artifact Removal):} By conditioning the physics encoder on an artifact-corrupted HSC image and the instrument encoder on a random set of HSC images, the model successfully removes severe instrumental distortions while preserving the underlying galaxy morphology.}
    \vspace{-10pt}
    \label{fig:data-driven-noise-model}
\end{figure}

\textbf{Data-driven noise model.} The instrument encoder and generative decoder together yield a data-driven noise model: by manipulating the instrument conditioning at inference time, we can simulate observations under different conditions while holding the underlying physics fixed. We demonstrate this in two complementary settings in Fig.~\ref{fig:data-driven-noise-model}. First, in a \textit{controlled SNR traversal}, we reconstruct an HSC target galaxy from its Legacy counterpart five times, fixing the initial noise sample $x_0$ but varying the instrument conditioning across the 5th, 25th, 50th, 75th, and 95th percentiles of the HSC SNR distribution (defined as the mean pixel-wise flux divided by the flux error, averaged across channels). Traversing this axis produces progressively cleaner reconstructions, even surpassing the apparent SNR of the ground truth---indicating an exploitable, manipulable noise model rather than memorization. Second, we use this noise model to \textit{remove strong systematics} from real observations: we pass an HSC image corrupted by severe instrumental distortion to the physics encoder and condition the instrument encoder on randomly selected HSC images to specify a baseline noise profile. Crucially, both encoders receive HSC images here (a combination never seen in training) yet the decoder reconstructs the underlying galaxy morphology while removing the artifacts, confirming that the model separates intrinsic physical signals from superimposed sensor distortions.

\newcamr{A complementary latent-swapping ablation supports this
separation: swapping the physics latent makes reconstructions
follow the donor galaxy, while swapping the instrument latent
makes their pixel-scale noise statistics follow the donor's
observing conditions
(Appendix~\ref{app:decoder_reliance},
Table~\ref{tab:decoder_conditioning_ablation}).}

\section{Discussion}

\textbf{Comparison with contrastive learning.} 
\newcamr{Contrastive learning across surveys encourages the encoding of information shared across observations of the same source. However, it does not explicitly encourage, and might even discourage, preserving fine detail visible only in a higher-resolution view. Such detail may nevertheless be scientifically informative. Instead, our method generates a softer pressure to align views and may help preserve more of that information.}
\newcamr{Moreover, contrastive learning does not natively support pixel-level reconstruction, while our counterfactual generative objective trains the representations to support pixel-level reconstruction.}

\newcamr{Appendix~\ref{app:contrastive} compares reconstruction quality, downstream property prediction performance, and the predictive power difference between lower and higher-resolution observations under the two training objectives. When used for counterfactual reconstruction, a decoder conditioned on frozen contrastive embeddings obtains worse reconstruction quality (25.5\% higher MSE) than our model, with a larger relative gap for the higher-resolution HSC Survey than for Legacy. When used for downstream prediction on physics-related properties from the physics latent space, our model outperforms contrastive learning in 6 out of 7 tasks. On the other hand, contrastive-learning produces instrument embeddings that achieve higher predictive power on all five evaluated instrument properties. Notably, the difference in predictive power between embeddings of the higher resolution and lower resolution surveys is greater with our model.}

\textbf{Training data requirements.}
\newcamr{To evaluate the data efficiency of our method, we vary the number of cross-matched training galaxies while keeping the instrument-neighbor pool fixed. With only 1,000 crossmatches (approximately 1\% of the full training set), the model recovers 81\% of the full-data improvement in mean $R^2$ for physics properties, but only 27--34\% of the performance for instrument properties (Appendix~\ref{app:data_efficiency}, Table~\ref{tab:data_efficiency}). These results suggest that useful physics representations can be learned with limited cross-survey overlap, while learning an instrument model benefits from access to large scale crossmatched data.}

\textbf{Limitations and future directions.} Our approach currently requires overlapping observations across instruments, which limits applicability to regions of shared sky coverage. In future work, we aim to extend the framework to unpaired settings where such overlap is unavailable or limited. Moreover, while we have focused on galaxy imaging, the framework is general and applicable to other domains in astrophysics and beyond.  In particular, we plan to apply it to the tens of millions of light curves from NASA's Transiting Exoplanet Survey Satellite \citep[TESS,][]{Ricker2015} and Kepler \citep{Borucki2010,Koch2010} missions.

\newcamr{Another limitation is that our experiments use bands with largely overlapping wavelength coverage across surveys and assume that matched views share information about the source, while measurement artifacts are weakly correlated across different instruments. Extending the framework to substantially different or non-overlapping spectral ranges remains untested. Such observations can probe distinct physical components of a source, so the model would need to distinguish wavelength-dependent source structure from instrumental effects.}

\section{Conclusion}

We have presented a framework for learning to disentangle instrumental effects from physical signals in scenarios with multiple views of the same object, i.e., observations from different instruments, using a simple counterfactual generative objective. Our dual-encoder architecture separates physics-invariant and instrument-specific factors structurally, without requiring hand-engineered contrastive losses or explicit instrument models. Compared to previous contrastive approaches, our method enables counterfactual generation by synthesizing how a source would appear under different observing conditions, and it is designed to scale to settings with more than two instruments. The generative component can also be viewed as a data-driven instrument model, learned implicitly from cross-matched observations. We have shown an application of our model to galaxy data in astrophysics, showing that the physics latent space recovers physical properties comparably to existing foundation models while being robust to instrument systematics, and that counterfactual generation produces realistic cross-survey translations.

\newpage
\begin{ack}

\newcamr{This work was supported in part by Advanced Micro Devices, Inc. under the AMD University Program's AI \& HPC Cluster. This work was supported by the MIT Generative AI Impact Consortium (MGAIC) and the MIT Office of Research Computing and Data (ORCD). The project that gave rise to these results received the support of a fellowship from ``la Caixa'' Foundation (ID 100010434). The fellowship code is B006068. We thank the AstroAI center at the Center for Astrophysics $\vert$ Harvard \& Smithsonian, where P~\mbox{M-P} was a summer researcher under the supervision of DM, Rafael Martinez-Galarza, and Cecilia Garraffo, for their support and useful discussions on multimodal foundation models for astronomy. We also thank Michael J. Smith for helping access the dataset, and Rocco Di Tella and Rebeka Bottger for insightful technical discussions.}
\end{ack}

\bibliography{iclr2026_conference}
\bibliographystyle{unsrtnat}  
\clearpage

\appendix
\section{Appendix}

\subsection{Model Architecture}\label{app:architecture}

We use a ResNet-18 \citep{he2015deepresiduallearningimage} architecture for the encoders, plus a \texttt{UNet2DConditionModel} \citep{von-platen-etal-2022-diffusers} for the flow-matching generative decoder. The final pooling and projection layers of the ResNet are replaced with a convolution that produces a $2\times2$ feature map with 16 channels, which is treated as a sequence of 4 tokens of size 16 and passed to the UNet via attention conditioning (matching the cross-attention dimension of 16). Table~\ref{tab:hpams1} shows the architecture parameters used in the final training run.

\newcamr{In our two-survey setting, the physics encoder provides 4 tokens from the same source observed by the other instrument, and the instrument encoder provides $4K$ tokens from $K$ neighboring sources observed by the target instrument. Concatenating these tokens along the sequence axis gives $4(1+K)$ conditioning tokens of dimension 16. For our training run we use five instrument neighbors so $K=5$ and we have a sequence of 24 tokens. This sequence is used as the key and value tokens for the cross-attention layers of the velocity network's cross-attention layers, while the UNet's spatial activations are the queries. Cross-attention is applied at every stage of the network, in all four downsampling blocks and all four upsampling blocks. Because cross-attention can intake a variable-length token sequence, the same decoder parameters can accommodate different numbers of conditioning views.}

\begin{table}[h]
\centering
\caption{Model architecture parameters for the conditional flow matching model with dual ResNet-18 encoders.}
\label{tab:hpams1} 
\begin{tabular}{ll}
\toprule
\textbf{Parameter} & \textbf{Value} \\
\midrule
\multicolumn{2}{l}{\textit{Data Dimensions}} \\
Image size & $48 \times 48$ \\
Input channels & 4 \\
Conditioning channels & 4 \\
\midrule
\multicolumn{2}{l}{\textit{UNet Architecture}} \\
Base model channels & 128 \\
Channel multiplier & $(1, 2, 4, 4)$ \\
Layers per block & 2 \\
Attention head dimension & 8 \\
Cross-attention dimension & 16 \\
Block types (down/up) & CrossAttnDownBlock2D / CrossAttnUpBlock2D \\
\midrule
\multicolumn{2}{l}{\textit{Encoder Architecture}} \\
Backbone & ResNet-18 (timm) \\
Pretrained weights & False \\
Projection layer & Replaced by Conv2d(512 $\rightarrow$ 16, kernel=1) \\
\bottomrule
\end{tabular}
\end{table}

\subsection{Training Configuration}

The model was trained using distributed data parallel across 4 NVIDIA H100 GPUs with \textit{bfloat16} mixed precision, completing 75,000 training steps in approximately 8 hours of wall-clock time. Table \ref{tab:hpam2} shows the training configuration and hyperparameters used in the final training run.

\begin{table}[h]
\centering
\caption{Training configuration and hyperparameters.}
\label{tab:hpam2}
\begin{tabular}{ll}
\toprule
\textbf{Parameter} & \textbf{Value} \\
\midrule
\multicolumn{2}{l}{\textit{Optimization}} \\
Optimizer & AdamW \\
Learning rate & $1 \times 10^{-4}$ \\
LR scheduler & CosineAnnealingLR \\
Loss function & Velocity MSE \\
\midrule
\multicolumn{2}{l}{\textit{Flow Matching}} \\
Integration steps (sampling) & 250 \\
Interpolation & $x_t = (1-t) x_0 + t x_1$ \\
Target velocity & $v(x_t, t, c) = x_1 - x_0$ \\
\midrule
\multicolumn{2}{l}{\textit{Training Details}} \\
Batch size & 64 \\
Total training steps & 75,000 \\
Number of devices & 4 (DDP)  \\
Precision & bf16-mixed \\
Validation check interval & Every 1,000 steps \\
Train/val split ratio & 95\% / 5\% \\
\bottomrule
\end{tabular}

\end{table}

\subsection{Model Used to Evaluate Downstream Performance} \label{app:MLP}
To evaluate the information content of our disentangled latent spaces, we perform supervised regression on a variety of physical and instrumental properties. We use a three-layer Multi-Layer Perceptron (MLP) with hidden dimensions of $[512, 256, 128]$, employing LayerNorm, GELU activations, and 20\% dropout. The models are optimized using a Smooth L1 loss to ensure robustness against outliers in the catalog labels.

\subsection{Galaxy Properties Used for Downstream Task}
\label{appendix:gal}

\paragraph{Dataset composition.}
All downstream-task evaluations are conducted on a single cross-matched sample
of $n = 5{,}469$ paired galaxies. The MLP probe is trained on a
random $90\%$ split of this sample and evaluated on the held-out $10\%$.

For the subset of physical properties that require DESI BGS spectroscopy
(redshift $z$, $\log M_\star$, sSFR, mass-weighted stellar age
$t_{\mathrm{age}}$ and mass-weighted metallicity $\log Z_{\mathrm{Met}}$, all
taken from the PROVABGS posterior catalogue), we drop the $\sim\!800$ rows
where the spectroscopic label is missing, leaving $n_{\mathrm{spec}} = 4{,}666$
galaxies. For every other target — ellipticity components $e_1,\,e_2$ (from
both the Legacy and HSC shape catalogues), Galactic extinction $E(B{-}V)$ on
the Legacy side and the HSC extinction $a$
(cross-matched in from the HSC catalogue
\texttt{a\_*} columns), as well as the instrumental metadata
(PSF size, PSF depth, galaxy depth and observation count) — the full
$n_{\mathrm{tot}} = 5{,}469$ sample is used. See below for a more in-depth description of the individual properties.

\paragraph{Physical properties.} Table~\ref{tab:physics_variables} summarizes the physical parameters used to evaluate the physics latent space. These span a range of galaxy properties: morphological quantities (ellipticity and half-light radius) measured directly from HSC imaging, spectroscopic redshift from DESI, and stellar population properties (stellar mass, specific star formation rate, metallicity, and stellar age) derived from Bayesian SED fitting by PROVABGS~\citep{Hahn2023}. The SED-derived quantities are inferred by jointly fitting stellar population synthesis models to DESI spectra and Legacy Survey photometry, and thus represent indirect measurements that integrate information beyond what is available in broadband imaging alone. Redshift is the most precisely determined of these quantities, as it is measured directly from spectral line positions.

\begin{table}[h]
\centering
\caption{Physical parameters used for downstream regression evaluation of the physics latent space.}
\label{tab:physics_variables}
\begin{tabular}{llp{7.5cm}}
\toprule
\textbf{Variable} & \textbf{Source} & \textbf{Description} \\
\midrule
Ellipticity ($e_1, e_2$) & HSC catalog & Galaxy shape ellipticity components, measured via the SDSS-style second moments of the surface brightness distribution. \\
Redshift ($z$) & DESI spectra & Spectroscopic redshift. \\
Stellar mass ($M_\star$) & PROVABGS & Total mass in stars,, derived from Bayesian spectral energy distribution (SED) fitting to DESI spectra~\citep{Hahn2023}. \\
Half-light radius ($R_\text{eff}$) & HSC catalog & Circularised radius enclosing half of the galaxy's total light. A measure of galaxy size. \\
sSFR & PROVABGS & Specific star formation rate $\log(\text{SFR}/M_\star)$: the rate at which new stars are forming normalised by the galaxy's stellar mass.  \\
$\log Z_\text{Met}$ & PROVABGS & Gas-phase metallicity: the abundance of elements heavier than helium in the galaxy's gas. Traces the galaxy's chemical enrichment history. \\
$t_\text{age}$ & PROVABGS & Mass-weighted stellar population age: the average age of the stars in the galaxy. \\
\bottomrule
\end{tabular}
\end{table}

\paragraph{Instrumental properties.} Table~\ref{tab:instrument_variables} summarizes the instrumental and observational parameters used to evaluate the instrument latent space. These quantities describe the local observing conditions at each galaxy's position and vary spatially across each survey's footprint. For the Legacy Survey, the catalog provides direct measurements of depth (both for point sources and extended galaxies), PSF size, number of contributing exposures, and foreground dust extinction. HSC provides analogous quantities for PSF size and dust extinction, though computed differently as noted below. We evaluate the instrument latent space on properties from both surveys to assess whether the encoder captures instrument-specific conditions for the survey it is given, without encoding information from the other survey.

\begin{table}[h]
\centering
\caption{Instrumental and observational parameters used for downstream regression evaluation of the instrument latent space. All quantities vary spatially across each survey's footprint.}

\label{tab:instrument_variables}
\begin{tabular}{llp{8cm}}
\toprule
\textbf{Variable} & \textbf{Bands} & \textbf{Description} \\
\midrule
\multicolumn{3}{l}{\textit{HSC (Hyper Suprime-Cam)}} \\
\midrule

\texttt{psf\_fwhm}        & $g, i, r, z$ & Effective size of the point spread function (arcsec): how much the instrument and atmosphere spread the light from a point source. Analogous to Legacy \texttt{PSFSIZE}.\textsuperscript{$\dagger$} \\

\texttt{a}                & $g, i, r, z$ & Milky Way dust extinction $A_\lambda$ per band (mag). Equivalent to Legacy \texttt{EBV} scaled by a filter-dependent coefficient: $A_\lambda = R_\lambda \, E(B-V)$.\textsuperscript{$\ddagger$} \\

\midrule
\multicolumn{3}{l}{\textit{Legacy Survey (DESI Legacy Imaging Surveys)}} \\
\midrule
\texttt{NOBS}      & $g, i, r, z$ & Number of exposures contributing to the coadd.\\

\texttt{GALDEPTH}  & $g, i, r, z$ & $5\sigma$ detection depth for a round exponential galaxy with $r_e = 0.45''$ (mag). Shallower than \texttt{PSFDEPTH} because galaxy flux is spread over more pixels. \\

\texttt{PSFDEPTH}  & $g, i, r, z$ & $5\sigma$ detection depth for a point source (mag). \\

\texttt{PSFSIZE}   & $g, i, r, z$ & Effective PSF size (arcsec). See HSC \texttt{psf\_fwhm} above.\textsuperscript{$\dagger$} \\

\texttt{EBV}       & --- & Milky Way dust extinction as colour excess $E(B-V)$ (mag). See HSC \texttt{a} above.\textsuperscript{$\ddagger$} \\
\bottomrule
\end{tabular}
\vspace{0.3em}

{\footnotesize $\dagger$ Both quantities measure how much the instrument and atmosphere spread the light from a point source, but are computed differently: HSC derives it from the second moments of the coadd PSF model at each source position~\citep{Aihara2018}; the Legacy Survey reports the exposure-weighted mean of the per-CCD PSF FWHM~\citep{Dey2019}.}

{\footnotesize $\ddagger$ Both are derived from the same Milky Way dust maps. Galactic extinction is a foreground astrophysical effect that depends on sky position, not on the instrument; it is included here because it modifies the observed flux analogously to an instrument systematic.}
\end{table}

\subsection{Downstream Task Results}
\label{app:downstream}

Table~\ref{tab:r2-scores} reports the full set of MLP-probe $R^2$ scores on the held-out \emph{overlap}-MMU sample, expanding on the summary in Fig.~\ref{fig:r2-scores}. We probe four representations: our physics latents (flattened across tokens), our instrument latents (mean-pooled across tokens), AION-1 (Base) embeddings~\citep{parker2025aion1omnimodalfoundationmodel}, and a frozen randomly initialized ResNet-18 with the same architecture as our encoders. For instrument properties only, we additionally include a \emph{cross-predict} baseline: a ResNet-18 trained from scratch to predict each survey's instrument properties from the \emph{other} survey's raw image. The two right-most columns are external reference values from the literature and are not compared against the other entries.

The physics latent space matches AION-1 within a few percentage points on most physics-related targets, despite using a much smaller latent dimension. The instrument latent space dominates on instrument-related properties, while the physics latents collapse near the random-baseline level on those same properties --- consistent with the disentanglement reported in the main text. The cross-predict baseline outperforms our instrument latents only on extinction-like quantities, where information about the other survey's footprint is sufficient to recover the target.

\textbf{The role of spatial-neighbor pairing.} Recall that the instrument encoder is conditioned on the five nearest spatial neighbors of the anchor within the same survey, motivated by the fact that nearby observations share similar instrument conditions. To validate this choice, we train an ablation variant in which the five conditioning galaxies are drawn at random instead. The instrument latent's recovery of instrument properties collapses (e.g. $R^2$ on Galaxy Depth drops from $0.68$ to $0.20$, and on $\#$ Observations from $0.50$ to nearly $0$), while physics-target performance changes only modestly. This confirms that spatially-close pairing supplies the training signal driving instrument-side disentanglement: random pairs do not present consistent enough instrument conditions for the model to separate them from physics.

\begin{table}[t]
  \centering
  \caption{MLP-probe $R^2$ on the held-out \emph{overlap}-MMU sample. \textbf{Bold} = best in row among the five probe methods (the two right-most columns are literature reference values and are not compared against). `Ours (Physics)' and `Ours (Instrument)' use our flow-matching encoder's $e_1$ (flat) and $e_2$ (mean-pooled) latents respectively. `AION-1 (Base)' is the AION-1 Base model~\citep{parker2025aion1omnimodalfoundationmodel} probed identically. `Rand.\ Init.\ ResNet' is a frozen randomly-initialised ResNet-18 of the same architecture as our encoders. `Cross-predict' is a fresh ResNet-18 trained from scratch to predict each survey's instrument properties from the \emph{other} survey's raw image (only defined for instrument targets). Reference values: AION-1-B (Ph+Im) from~\citep{parker2025aion1omnimodalfoundationmodel} Table~1; AstroCLIP image-only on Legacy $\{g,r,z\}$ from~\citep{lanusse2023astroclip}. Band-averaged where applicable.}
  \label{tab:r2-scores}
  \setlength{\tabcolsep}{3pt}
  \resizebox{\linewidth}{!}{%
  \begin{tabular}{l rrrrr cc}
    \toprule
    \multirow{2}{*}{Property} & \multicolumn{5}{c}{Probe on cross-matched Legacy-HSC sample (Im)} & \multicolumn{2}{c}{Literature Reported Values} \\
    \cmidrule(lr){2-6} \cmidrule(lr){7-8}
    & Ours (Phys) & Ours (Instr) & AION-1-B & Rnd In. ResNet & X-Predict & AION-1-B \,(Ph+Im) & AstroCLIP \,(Im) \\
    \midrule
    \multicolumn{8}{l}{\textit{Phys.-related properties}} \\
    Ellipticity $e_2$ & 0.934 & 0.321 & \textbf{0.941} & 0.616 & — & — & — \\
    Ellipticity $e_1$ & 0.918 & 0.421 & \textbf{0.939} & 0.613 & — & — & — \\
    Redshift $z$ & 0.801 & 0.535 & \textbf{0.857} & 0.514 & — & 0.930 & 0.780 \\
    $\log M_\star$ & 0.755 & 0.491 & \textbf{0.818} & 0.485 & — & 0.890 & 0.730 \\
    sSFR & 0.558 & 0.297 & \textbf{0.624} & 0.312 & — & 0.640 & 0.420 \\
    $t_{\mathrm{age}}$ & 0.257 & 0.114 & \textbf{0.353} & 0.091 & — & 0.450 & 0.290 \\
    $\log Z_{\mathrm{Met}}$ & 0.250 & 0.155 & \textbf{0.276} & 0.123 & — & 0.490 & 0.360 \\
    \midrule
    \multicolumn{8}{l}{\textit{Legacy Ins. Properties}} \\
    Galaxy Depth & 0.185 & \textbf{0.682} & 0.426 & 0.202 & 0.214 & — & — \\
    PSF Depth & 0.181 & \textbf{0.667} & 0.409 & 0.191 & 0.210 & — & — \\
    PSF size & 0.284 & \textbf{0.556} & 0.290 & 0.111 & 0.313 & — & — \\
    Extinction $E(B{-}V)$ & 0.159 & 0.551 & 0.326 & 0.232 & \textbf{0.648} & — & — \\
    \# Observations & 0.074 & \textbf{0.501} & 0.166 & 0.073 & 0.197 & — & — \\
    \midrule
    \multicolumn{8}{l}{\textit{HSC Ins. Properties}} \\
    Extinction $a$ & 0.114 & 0.546 & 0.299 & 0.204 & \textbf{0.669} & — & — \\
    PSF size & 0.069 & \textbf{0.438} & 0.097 & 0.021 & 0.368 & — & — \\
    \bottomrule
  \end{tabular}%
  }
\end{table}

\subsection{Generative Model Evaluation}
\label{app:generative-eval}

\begin{figure}[t]
    \centering
    \includegraphics[width=0.8\linewidth]{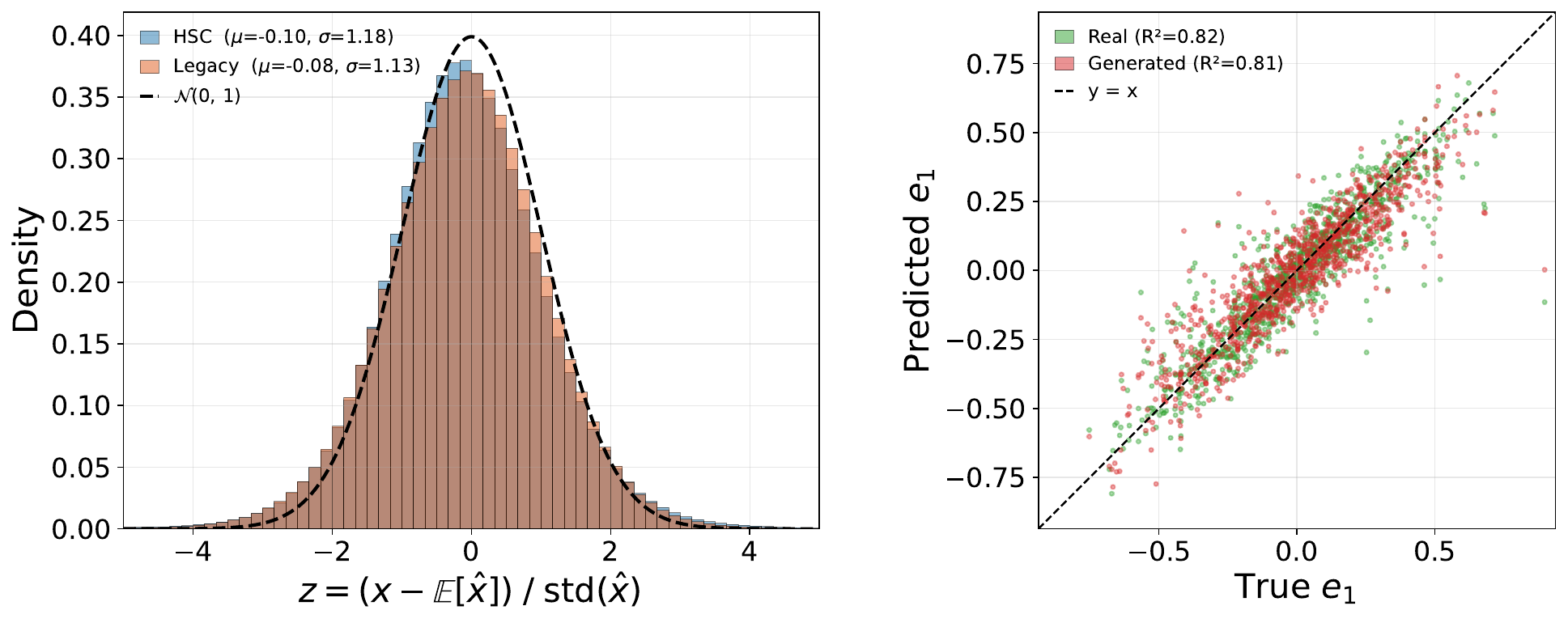}
    \caption{\textbf{Left:} Pixel-wise Z-score distribution of generated posterior samples relative to ground-truth target images, across all pixels and held-out galaxies. Both HSC-anchored and Legacy-anchored reconstructions closely approximate a standard normal distribution, indicating that the posterior mean is approximately unbiased, the posterior variance captures the true prediction error, and that the residual uncertainty is approximately Gaussian. The standard deviations slightly above unity indicate mild overconfidence, more pronounced for HSC-anchored reconstructions, consistent with the greater ambiguity in generating higher-resolution images from lower-resolution conditioning. \textbf{Right:} $R^2$ for galaxy ellipticity from a ResNet trained on real HSC images, evaluated on real versus counterfactual HSC images. Near-identical performance shows that existing HSC pipelines can be readily applied to Legacy images through our counterfactual model, extending HSC-quality analysis across the full Legacy footprint.}
    \label{fig:generative_metrics}
\end{figure}

We assess the counterfactual generative decoder along three complementary axes: (1) preservation of spatial structure across scales, (2) calibration of the pixel-wise posterior, and (3) preservation of intrinsic galaxy morphology under pipeline transfer (the last is presented in the main text; see also Fig.~\ref{fig:generative_metrics}, right). All evaluations use a held-out test set, with 32 posterior samples generated per anchor.

\begin{figure}[ht]
    \centering
    \includegraphics[width=0.99\linewidth]{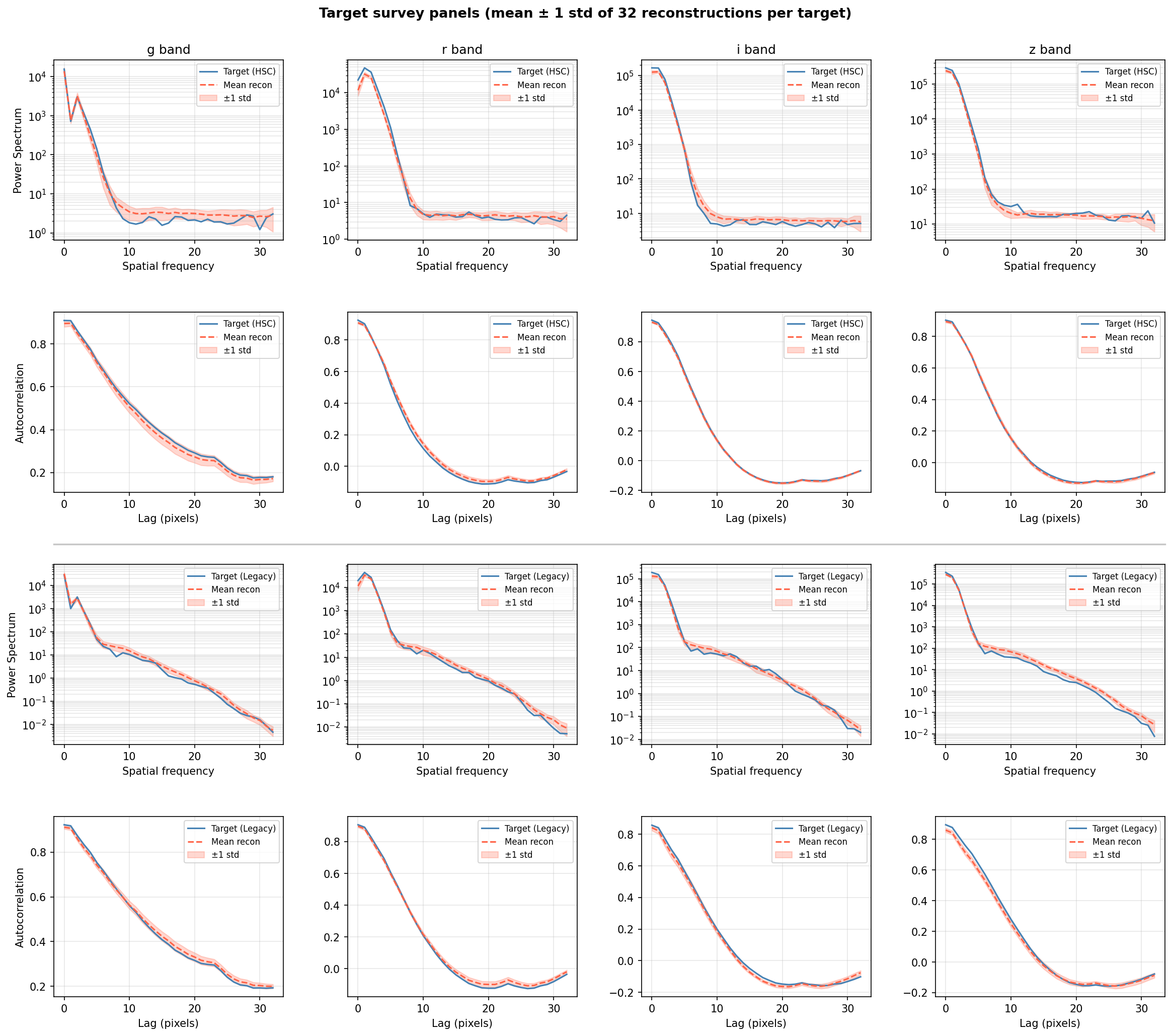}
    \caption{\textbf{Spatial Structure Preservation.} Power spectrum (left) and autocorrelation (right) per band for ground-truth and generated images. Each plot shows the true value (blue), the posterior mean from 32 generated samples (orange dashed line), and the $1\sigma$ posterior standard deviation (orange shaded region). Strong agreement across spatial frequencies and pixel separation (lag) distances indicates that the model captures realistic morphological structure and noise properties.}
    \label{fig:power_autocorr}
\end{figure}

\textbf{Spatial structure preservation.} To test that the generated images preserve realistic spatial structure and correlation patterns, we compute the power spectral density (PSD) and pixel-lag autocorrelation for each of the four imaging bands, in both reconstruction directions (Legacy$\to$HSC and HSC$\to$Legacy). Fig.~\ref{fig:power_autocorr} shows strong agreement between ground truth and generated samples across spatial scales for both directions, indicating that the model reproduces realistic galaxy morphologies and instrument noise characteristics rather than pixel-level artifacts (which would manifest as excess high-frequency power) or overly-smoothed outputs (which would appear as a deficit). The posterior uncertainty bands (shaded regions) appropriately expand at higher spatial frequencies, where noise dominates.

\textbf{Pixel-wise posterior calibration.} We further validate the uncertainty calibration of the flow-matching posterior using the pixel-wise Z-score,
\begin{equation*}
    Z = \frac{x - \mathbb{E}[\hat{x}]}{\mathrm{std}(\hat{x})},
\end{equation*}
where $x$ is the ground-truth pixel value and $\mathbb{E}[\hat{x}]$, $\mathrm{std}(\hat{x})$ are the posterior mean and standard deviation estimated from the generated samples. Fig.~\ref{fig:generative_metrics} (left) shows that $Z$ closely follows a standard normal for both survey directions, indicating that the posterior mean is approximately unbiased and the residual uncertainty is approximately Gaussian. The standard deviations slightly above unity indicate mild overconfidence (the model underestimates the true pixel-wise variance by ${\sim}15\%$) and the effect is more pronounced for HSC-anchored reconstructions, consistent with the greater ambiguity in generating higher-resolution images from lower-resolution conditioning.

\subsection{Normalizing Flow for Outlier Detection}
\label{app:flow}

To rank candidate outliers, we fit a normalizing flow to the empirical density of the embeddings in each latent space and use the negative log-likelihood $-\log p(z)$ as the anomaly score. We use a Neural Spline Flow (NSF; \citealp{durkan2019neuralsplineflows}) implemented in \texttt{zuko}~\citep{rozet2023zuko}, with 6 coupling transforms and a 2-layer MLP of width 64 in each transform. The flow is unconditional (\texttt{context=0}) and is trained by maximum likelihood with Adam.

We use the embeddings from just the HSC images for this experiment. We fit a separate flow for each representation we score: (i) our physics latent flattened across the 4 spatial tokens (64D), (ii) our instrument latent mean-pooled across tokens (16D), and (iii) AION-1 (Base) 768D mean-pooled embedding. For AION-1 we also tried fitting the flow to embeddings reduced to 64D via PCA fit on the training split and obtained qualitatively similar outliers. For each space we use 80\% of the $\sim\!100{,}000$ HSC galaxies to fit the flow and score the full sample under the trained model; the lowest-likelihood examples are the candidate outliers shown in Fig.~\ref{fig:anomalies}. Table~\ref{tab:flow_hpams} lists the hyperparameters used.

\begin{table}[h]
\centering
\caption{Normalizing flow architecture and training hyperparameters used for outlier detection.}
\label{tab:flow_hpams}
\begin{tabular}{ll}
\toprule
\textbf{Parameter} & \textbf{Value} \\
\midrule
\multicolumn{2}{l}{\textit{Flow Architecture}} \\
Family & Neural Spline Flow (NSF, \texttt{zuko}) \\
Number of transforms & 6 \\
Hidden features per transform & $[64, 64]$ \\
Context dimension & 0 (unconditional) \\
Input dimension & 64 / 16 (Ours) / 768 (AION-1) \\
\midrule
\multicolumn{2}{l}{\textit{Optimization}} \\
Loss & Negative log-likelihood \\
Optimizer & Adam \\
Learning rate & $1 \times 10^{-3}$ \\
Batch size & 512 \\
Epochs & 50 \\
Model selection & Lowest mean training NLL (best-state checkpoint) \\
\midrule
\multicolumn{2}{l}{\textit{Data Split}} \\
Train fraction & 80\% \\
Scoring set & All $\sim\!100{,}000$ HSC galaxies \\
Random seed & 42 \\
\bottomrule
\end{tabular}
\end{table}

\newpage

\subsection{Decoder Reliance on Physics and Instrument Conditioning}
\label{app:decoder_reliance}

\newcamr{We perform an experiment to further test whether the decoder
actually uses the two conditioning latents. For 256 held-out
galaxies, we generate samples from the same initial noise
realization $x_0$ under three conditions: ``intact'' conditioning
(with the physics latent $z_{\mathrm{phy}}$ obtained from the same
galaxy observed by the other survey, and the instrument latent
$z_{\mathrm{ins}}$ obtained from the anchor's spatial neighbors in
the target survey), $z_{\mathrm{phy}}$ replaced by that of a
different held-out galaxy (which we will refer to as the
``donor''), and $z_{\mathrm{ins}}$ replaced by the donor's
instrument conditioning. Donors are drawn within the target
survey, so the intervention is not a survey swap. Crucially,
each donor also has a ground truth pair in the target survey, so we can
test not only how the swap affects agreement with the target but
also how the generation follows the donor. We average over 4
posterior samples per galaxy, test both surveys as targets, and
report the results in
Table~\ref{tab:decoder_conditioning_ablation}.}

\newcamr{To assess the dependence on physics conditioning, we use the
reconstruction MSE as a proxy for source identity.} \newcamr{To assess the dependence on instrument conditioning, we define
$\sigma_{\mathrm{ap}}$ as a robust, MAD-based estimate of the
standard deviation of adjacent-pixel differences, i.e.\ a proxy
for the pixel-scale noise of an image. For each galaxy and conditioning setup, we have three values:
$\sigma_{\mathrm{ap},\mathrm{gen}}$ for the generated image,
$\sigma_{\mathrm{ap},\mathrm{own}}$ for the real anchor observation,
and $\sigma_{\mathrm{ap},\mathrm{donor}}$ for the donor's observation
in the target survey.}

\newcamr{We evaluate the correlation $\rho$ between these quantities
across all the evaluated galaxies, separately for each target
survey and condition, to test whose observing
conditions the generation inherits. Here,
$\rho(\sigma_{\mathrm{ap},\mathrm{gen}},
      \sigma_{\mathrm{ap},\mathrm{own}})\approx1$ means that the
generated noise level tracks its own target's as expected when running the model normally,
$\rho(\sigma_{\mathrm{ap},\mathrm{gen}},
      \sigma_{\mathrm{ap},\mathrm{donor}})\approx1$ means that it
tracks the donor's, and $\rho\approx0$ means little correlation
across galaxies.}

\begin{table}[h]
\centering
\small
\setlength{\tabcolsep}{4pt}
\renewcommand{\arraystretch}{1.15}
\caption{\newcamr{Decoder use of physics and instrument conditioning.
We compare intact conditioning with replacing either latent by
that of a donor galaxy within the same target survey.
MSE measures agreement with the real anchor (own) or donor
observation. $\rho$ measures how the generated image's
pixel-scale noise correlates with
the anchor's or donor's across galaxies.
HSC and Legacy denote the target survey.
Bold entries highlight agreement with the donor after the
corresponding swap.
}}
\label{tab:decoder_conditioning_ablation}
\newcamr{
\begin{tabular}{@{}lrrrrrr@{}}
\toprule
& \multicolumn{2}{c}{Intact}
& \multicolumn{2}{c}{Physics swapped}
& \multicolumn{2}{c}{Instrument swapped} \\
\cmidrule(lr){2-3}
\cmidrule(lr){4-5}
\cmidrule(lr){6-7}
Metric & HSC & Legacy & HSC & Legacy & HSC & Legacy \\
\midrule
MSE vs.\ own target
& 0.058 & 0.186
& 0.970 & 1.490
& 0.060 & 0.186 \\
MSE vs.\ donor target
& 0.970 & 1.480
& \textbf{0.058} & \textbf{0.188}
& 0.970 & 1.480 \\
\midrule
$\rho(\sigma_{\mathrm{ap},\mathrm{gen}},
       \sigma_{\mathrm{ap},\mathrm{own}})$
& $+0.99$ & $+0.94$
& $+0.90$ & $+0.72$
& $-0.03$ & $+0.17$ \\
$\rho(\sigma_{\mathrm{ap},\mathrm{gen}},
       \sigma_{\mathrm{ap},\mathrm{donor}})$
& $-0.11$ & $-0.10$
& $-0.01$ & $+0.10$
& $\mathbf{+0.89}$ & $\mathbf{+0.69}$ \\
\bottomrule
\end{tabular}
}
\end{table}

\newcamr{Shuffling $z_{\mathrm{phy}}$ changes the image to track the donor galaxy, as shown by the MSE against the donor's ground truth image closely matching the intact model's MSE against its own target (0.058/0.188 vs 0.058/0.186), while the noise statistics stay consistent with those of the original target (correlation goes from 0.99 / 0.94 to 0.90 / 0.72). This means our reconstruction matches the large-scale structure (mostly driven by source identity) of the donor galaxy, while the small scale noise structure continues to match that of the original target.}

\newcamr{Shuffling $z_{\mathrm{ins}}$ has the opposite effect: source identity is untouched (the MSE difference is $0.002\pm0.002 / 0.000\pm0.005$, a null result), while the generated noise is different from the original target's instrument properties and tracks that of the donor ( $\rho = -0.03/+0.17$ for the original target, and  $\rho = +0.89/+0.69$  on the donor).}





\newpage
\subsection{Ablation: Counterfactual Generative Objective vs. Contrastive Learning}\label{app:contrastive}

\newcamr{To evaluate the role of the training objective, we train the same physics and instrument encoders with an InfoNCE contrastive loss \citep{oord2018representation} instead of
our counterfactual generative objective.}

\newcamr{For the physics encoder, positives pairs consist of an anchor galaxy image and its cross-survey pair, while negative pairs combine the anchor with images of other galaxies coming from a mixture of both surveys (all the other anchor images in the batch).}

\newcamr{For the instrument encoder, positive pairs consist of an anchor and its sky neighbours observed by the same instrument, while negative pairs combine the anchor with neighbors of other anchors in the same batch. Following \citet{audenaert2025causalfoundationmodelsdisentangling}, we attach SimCLR projection heads \citep{chen2020simclr} to both encoders and compute the contrastive losses on the projected embeddings.}

\newcamr{Both models use the same encoder architectures and
training dataset. A notable difference is that the contrastive baseline must pass the anchor image through the encoders to construct the training pairs, while in our model the anchor image is never an input to the encoders. All 5,469 galaxies used for downstream evaluation are excluded
from the self-supervised training stage for both models.}

\newcamr{We compare reconstruction quality and downstream predictive performance, and examine how the predictive power of models trained with each training objective respond to the addition of higher-resolution observations.}


\newcamr{\textbf{Reconstruction.} We train a flow-matching decoder conditioned on frozen contrastive embeddings, using the same counterfactual triplet reconstruction logic as our model. This proves how well the learned representations support counterfactual reconstruction. Table~\ref{tab:contrastive_reconstruction} reports reconstruction
MSE on 512 held-out validation galaxies.}

\begin{table}[h]
\centering
\small
\setlength{\tabcolsep}{4pt}
\renewcommand{\arraystretch}{1.15}
\caption{\newcamr{Counterfactual reconstruction MSE for our model
and a decoder conditioned on frozen contrastive embeddings.
Relative increases are measured against our model.}}
\label{tab:contrastive_reconstruction}
\newcamr{
\begin{tabular}{@{}lccc@{}}
\toprule
\textbf{Conditioning style}
& \textbf{MSE HSC}
& \textbf{MSE Legacy}
& \textbf{MSE all} \\
\midrule
Ours
& $0.0474 \pm 0.0031$
& $0.1610 \pm 0.0047$
& $0.1042$ \\
Frozen contrastive embeddings
& $0.0679 \pm 0.0045$
& $0.1936 \pm 0.0071$
& $0.1308$ \\
\midrule
Relative MSE increase
& \textbf{+43.1\%}
& \textbf{+20.3\%}
& \textbf{+25.5\%} \\
\bottomrule
\end{tabular}
}
\end{table}

\newcamr{Frozen contrastive embeddings obtain a reconstruction MSE 25.5\% greater than our model. The relative increase is larger for HSC (43.1\%) than for Legacy (20.3\%). These results suggest that our generative objective learns representations better suited for counterfactual reconstruction, and that the largest relative benefit is for the higher-resolution targets, indicating that it does not suffer as much from the information asymmetry between the two surveys.}


\newcamr{\textbf{Downstream $\mathbf{R^2}$.} We freeze both models and fit an MLP probe on each latent, using five-fold cross-validation over the 5,469 evaluation galaxies. Our physics latent achieves higher $R^2$ than the contrastive one on 6 of the 7 physics and morphology properties. For log-metallicity $logZ$, the two agree within the fold-to-fold scatter. Nonetheless, the contrastive instrument latents predict all five instrument properties better than ours, yet its physics latents retain even less information about instrument properties than ours does (Table~\ref{tab:contrastive_downstream}).}

\newcamr{We attribute this to what each objective requires. The contrastive loss punishes anything the anchor shares with its negatives. Dust extinction-induced color excess E(B-V) is a good example: nearby galaxies share it, so it gets pushed out of the physics latent even though both physics views have it, and pulled into the instrument latent, whose negatives come from other parts of the sky. Our proposed objective has no such pressure, so this information can be partially encoded in the physics latent space and these factors settle wherever the decoder needs them. Our physics latent space holds shared information, but not instrument-private information. The contrastive objective latent space discards shared information, and that costs it at generation time (Table~\ref{tab:contrastive_reconstruction}).}

\begin{table}[h]
\centering
\small
\setlength{\tabcolsep}{4pt}
\caption{Downstream $R^2$ ($\times 100$) for embeddings trained using our counterfactual generative objective versus those trained using a contrastive learning baseline. (a)~Each property is predicted from its matching latent: physics-related properties (left) from the physics latent and instrument properties (right) from the instrument latent. (b)~Each property is predicted from the \emph{other} latent, measuring how much of it is carried across the split.}
\label{tab:contrastive_downstream}

\begin{tabular}{@{}lccc@{\hspace{8pt}}lccc@{}}
\toprule
\multicolumn{4}{@{}c}{\textbf{Physics-related properties}}
& \multicolumn{4}{c@{}}{\textbf{Instrument properties}} \\
\cmidrule(r{6pt}){1-4}\cmidrule(l){5-8}
Property & Ours & Contrastive & $\Delta$
& Property & Ours & Contrastive & $\Delta$ \\
\midrule
\multicolumn{4}{@{}l}{\emph{(a) from the physics latent}}
& \multicolumn{4}{l@{}}{\emph{(a) from the instrument latent}} \\
$z$         & $\boldsymbol{77.0 \pm 1.3}$ & $75.4 \pm 1.5$ & $+1.5 \pm 0.3$
& Galaxy depth & $75.1 \pm 1.1$ & $\boldsymbol{85.2 \pm 1.6}$ & $-10.0 \pm 0.2$ \\
$\log M_\star$ & $\boldsymbol{73.0 \pm 0.8}$ & $69.6 \pm 1.2$ & $+3.4 \pm 0.6$
& PSF depth    & $74.2 \pm 1.3$ & $\boldsymbol{84.6 \pm 2.0}$ & $-10.4 \pm 0.3$ \\
sSFR        & $\boldsymbol{51.0 \pm 1.4}$ & $48.4 \pm 1.9$ & $+2.5 \pm 0.7$
& $N_{\mathrm{obs}}$ & $53.5 \pm 1.4$ & $\boldsymbol{69.9 \pm 0.9}$ & $-16.4 \pm 0.7$ \\
$t_{\mathrm{age}}$ & $\boldsymbol{22.3 \pm 2.7}$ & $16.5 \pm 2.0$ & $+5.8 \pm 0.9$
& $E(B{-}V)$   & $59.1 \pm 2.5$ & $\boldsymbol{77.0 \pm 1.7}$ & $-17.9 \pm 0.9$ \\
$\log Z$    & $18.5 \pm 3.3$ & $19.1 \pm 1.0$ & $-0.6 \pm 1.4$
& HSC PSF FWHM & $50.0 \pm 1.7$ & $\boldsymbol{68.4 \pm 1.8}$ & $-18.4 \pm 0.6$ \\
$e_1$       & $\boldsymbol{94.3 \pm 0.7}$ & $89.9 \pm 1.1$ & $+4.4 \pm 0.4$
& & & & \\
$e_2$       & $\boldsymbol{94.9 \pm 0.7}$ & $90.9 \pm 1.3$ & $+4.0 \pm 0.4$
& & & & \\
\midrule
\multicolumn{4}{@{}l}{\emph{(b) from the instrument latent}}
& \multicolumn{4}{l@{}}{\emph{(b) from the physics latent}} \\
$z$         & $63.4 \pm 2.7$ & $50.4 \pm 1.9$ & $+13.0 \pm 1.6$
& Galaxy depth & $17.0 \pm 1.0$ & $-0.8 \pm 0.6$ & $+17.8 \pm 0.3$ \\
$\log M_\star$ & $61.0 \pm 1.3$ & $49.8 \pm 2.5$ & $+11.2 \pm 1.4$
& PSF depth    & $16.6 \pm 0.9$ & $-0.6 \pm 0.7$ & $+17.2 \pm 0.2$ \\
sSFR        & $41.7 \pm 2.5$ & $32.3 \pm 1.4$ & $+9.4 \pm 0.9$
& $N_{\mathrm{obs}}$ & $5.9 \pm 1.3$ & $-2.5 \pm 0.7$ & $+8.4 \pm 0.3$ \\
$t_{\mathrm{age}}$ & $11.3 \pm 1.0$ & $5.2 \pm 1.9$ & $+6.1 \pm 0.7$
& $E(B{-}V)$   & $17.6 \pm 2.0$ & $6.6 \pm 2.0$ & $+11.0 \pm 1.0$ \\
$\log Z$    & $17.7 \pm 2.0$ & $6.1 \pm 1.9$ & $+11.6 \pm 1.1$
& HSC PSF FWHM & $7.0 \pm 1.1$ & $2.4 \pm 0.8$ & $+4.6 \pm 0.3$ \\
$e_1$       & $59.2 \pm 1.2$ & $39.4 \pm 3.1$ & $+19.8 \pm 1.7$
& & & & \\
$e_2$       & $70.7 \pm 3.1$ & $49.8 \pm 4.1$ & $+20.9 \pm 1.4$
& & & & \\
\bottomrule
\end{tabular}
\end{table}

\newpage
\newcamr{\textbf{Predictive advantage of higher-resolution observations.}
To examine whether the learned representations retain fine detail visible only in the higher-resolution HSC survey, we compare downstream property prediction using physics latents extracted separately from HSC and Legacy images of the same galaxies.
For each model and property, we compute
$\Delta_{\mathrm{survey}} = R^2(z_{\mathrm{phy,HSC}}) - R^2(z_{\mathrm{phy,Legacy}})$,
where $z_{\mathrm{phy,HSC}}$ and $z_{\mathrm{phy,Legacy}}$ are obtained by encoding only the HSC or Legacy image, respectively, and $R^2(z)$ denotes downstream prediction performance using
features $z$. 
Thus, $\Delta_\mathrm{survey}$ measures how much more information about a property is recoverable from the sharper survey's view than from the shallower one's. An objective that keeps only what the two views share should thus give $\Delta_\mathrm{survey} \approx 0$.}


\begin{table}[h]
\centering
\small
\setlength{\tabcolsep}{4pt}
\renewcommand{\arraystretch}{1.15}
\caption{\newcamr{Survey differences in downstream property-prediction $R^2$. For each property,
$\Delta_{\mathrm{survey}}$ is the HSC-only score minus the Legacy-only score using physics latents. The final column reports the difference between models (ours minus contrastive).}}
\label{tab:contrastive_survey_gap}
\newcamr{
\begin{tabular*}{\linewidth}{@{\extracolsep{\fill}}lrrr@{}}
\toprule
\textbf{Property}
& \textbf{Ours}
& \textbf{Contrastive}
& \textbf{Difference} \\
\midrule
Redshift $z$
& $+0.015 \pm 0.002$ & $+0.005 \pm 0.003$ & $+0.009 \pm 0.004$ \\
$\log M_\star$
& $+0.028 \pm 0.009$ & $+0.005 \pm 0.006$ & $+0.023 \pm 0.010$ \\
sSFR
& $+0.013 \pm 0.004$ & $-0.002 \pm 0.003$ & $+0.015 \pm 0.005$ \\
Stellar age
& $+0.005 \pm 0.014$ & $-0.007 \pm 0.003$ & $+0.011 \pm 0.014$ \\
$\log Z$
& $+0.010 \pm 0.003$ & $+0.007 \pm 0.004$ & $+0.003 \pm 0.005$ \\
Ellipticity $e_1$
& $+0.026 \pm 0.003$ & $+0.007 \pm 0.002$ & $+0.019 \pm 0.004$ \\
Ellipticity $e_2$
& $+0.019 \pm 0.004$ & $+0.010 \pm 0.002$ & $+0.009 \pm 0.003$ \\
\bottomrule
\end{tabular*}
}
\end{table}


\newcamr{Table~\ref{tab:contrastive_survey_gap} shows a larger HSC advantage for our model than for the contrastive baseline on all seven evaluated properties. The contrastive baseline is close to $\Delta_\mathrm{survey} \approx 0$, meaning its HSC and Legacy latents are close to interchangeable for the purposes of inference, which is what a collapse toward the lower-resolution view looks like, while ours retains  $\Delta_\mathrm{survey}>0$. We expect this advantage to grow as the resolution gap widens: for example, when combining ground- and space-based observations, where forcing agreement between views risks discarding precisely the fine-scale information of greatest scientific value.}

\newpage

\subsection{Data Efficiency: Scaling with Cross-Matched Training Set Size}\label{app:data_efficiency}



\newcamr{
When applying our method to new surveys, the availability of cross-matched observations is the main constraint, not the amount of single survey data.
Training requires observations of the same galaxy from both instruments to form the anchor and physics pair views. In contrast, the same-instrument conditioning neighbors need not have a counterpart in the other survey.  For example, the model in the main text trains on ${\sim}100{,}000$ crossmatched anchors but draws its neighbours from a pool of ${\sim}468{,}000$ sources, most of which have no crossmatch. 

To assess how performance depends on the available overlap, we train models on nested subsets of 1,000, 3,162, 10,000, and 31,622 cross-matched galaxies, while keeping the instrument-neighbor pool fixed at ${\sim}468{,}000$ sources (Table~\ref{tab:data_efficiency}). This is the most realistic deployment scenario for a new survey, which might have a large pool of data but only limited overlap with previous surveys. }

\newcamr{Because an untrained encoder already provides nonzero
predictive performance, we measure the percentage of the
full-data improvement above this baseline recovered at each
training set size. For each property group, we define the fraction of full-data improvement recovered as}

\begin{equation*}
    \eta(N)=\frac{\bar{R}^2_N-\bar{R}^2_\mathrm{untrained}}{\bar{R}^2_{N_\mathrm{max}}-\bar{R}^2_\mathrm{untrained}}
\end{equation*}

\newcamr{where $\bar{R}^2_N$ is the mean downstream score using $N$ training crossmatches and $\bar{R}^2_{N_\mathrm{max}}$ is the full-data reference. This assigns 0\% to the untrained encoder and 100\% to the checkpoint reported in the main text}.

\begin{table}[h]
\centering
\small
\setlength{\tabcolsep}{4pt}
\renewcommand{\arraystretch}{1.15}
\caption{Downstream performance as a function of
number of crossmatches in the training set. Columns show the mean $R^2$ across seven physics, five Legacy instrument, and two HSC instrument properties, with the percentage of improvement recovered in parentheses. The full-data model is the model used in the main text, its training split contained approximately 99,000 crossmatches. The instrument-neighbor pool is kept fixed across training subsets.}
\label{tab:data_efficiency}
\begin{tabular*}{\linewidth}{@{\extracolsep{\fill}}l
  r@{\extracolsep{0pt}\,}r
  @{\extracolsep{\fill}\quad}r@{\extracolsep{0pt}\,}r
  @{\extracolsep{\fill}\quad}r@{\extracolsep{0pt}\,}r@{}}
\toprule
\shortstack[l]{\textbf{Training}\\\textbf{crossmatches}}
& \multicolumn{2}{c}{\shortstack{\textbf{Physics}\\(7 properties)}}
& \multicolumn{2}{c}{\shortstack{\textbf{Legacy instrument}\\(5 properties)}}
& \multicolumn{2}{c@{}}{\shortstack{\textbf{HSC instrument}\\(2 properties)}} \\
\midrule
1,000
& 0.593 & (81.2\%) & 0.289 & (33.8\%) & 0.219 & (26.8\%) \\
3,162
& 0.612 & (89.0\%) & 0.390 & (55.9\%) & 0.282 & (43.7\%) \\
10,000
& 0.625 & (94.3\%) & 0.471 & (73.7\%) & 0.353 & (62.7\%) \\
31,622
& 0.634 & (98.0\%) & 0.499 & (79.8\%) & 0.375 & (68.6\%) \\
\midrule
Full-data model
& 0.639 & (100.0\%) & 0.591 & (100.0\%) & 0.492 & (100.0\%) \\
Untrained encoder
& 0.394 & (0.0\%) & 0.135 & (0.0\%) & 0.119 & (0.0\%) \\
\bottomrule
\end{tabular*}
\end{table}

\newcamr{Table~\ref{tab:data_efficiency} shows that with just 1,000 crossmatches (approximately 1\% of the full training set) the model recovers 81.2\% of the full-data improvement in mean $R^2$ for physics properties, but only 27-34\% for instrument properties. At 31,622 crossmatches, these fractions reach 98.0\% and 69-80\%, respectively. These results suggest that most of the performance in physics-related tasks can be recovered with as little as $10^3$ cross-survey matches, while prediction of instrument properties continues to improve across the studied range all the way to $10^5$.}

\clearpage

\section*{NeurIPS Paper Checklist}

\begin{enumerate}

\item {\bf Claims}
    \item[] Question: Do the main claims made in the abstract and introduction accurately reflect the paper's contributions and scope?
    \item[] Answer: \answerYes{}
    \item[] Justification: The abstract and introduction state our three main contributions: a counterfactual generative objective for disentangling physics from instrument factors, and the resulting downstream capabilities (parameter inference, outlier detection, counterfactual generation, instrument-independent retrieval, and zero-shot pipeline transfer). Each is supported by the experiments described in Section~\ref{sec:applicationtoastro} and the appendices.

\item {\bf Limitations}
    \item[] Question: Does the paper discuss the limitations of the work performed by the authors?
    \item[] Answer: \answerYes{}
    \item[] Justification: A dedicated ``Limitations and future directions'' paragraph appears at the end of the main text, noting that our framework requires overlapping observations across instruments. We also acknowledge that the instrument latent retains residual physics information that we cannot quantify, and in Appendix~\ref{app:generative-eval} we mention that the posterior pixel-wise variance is mildly underestimated ($\sim 15\%$).

\item {\bf Theory assumptions and proofs}
    \item[] Question: For each theoretical result, does the paper provide the full set of assumptions and a complete (and correct) proof?
    \item[] Answer: \answerNA{}
    \item[] Justification: The paper does not present formal theoretical results; the contributions are methodological and empirical.

    \item {\bf Experimental result reproducibility}
    \item[] Question: Does the paper fully disclose all the information needed to reproduce the main experimental results of the paper to the extent that it affects the main claims and/or conclusions of the paper (regardless of whether the code and data are provided or not)?
    \item[] Answer: \answerYes{}
    \item[] Justification: The methods section describes the dual-encoder architecture, counterfactual flow-matching objective, triplet construction, and spatial-neighbor pairing strategy. Appendix~\ref{app:architecture} details the modified ResNet-18 encoder design and the UNet velocity network, and Appendix (Training Configuration) lists hyperparameters, optimizer, schedule, and number of training steps. The cross-matched Legacy--HSC dataset is constructed from publicly available surveys, and the preprocessing pipeline is fully described.

\item {\bf Open access to data and code}
    \item[] Question: Does the paper provide open access to the data and code, with sufficient instructions to faithfully reproduce the main experimental results, as described in supplemental material?
    \item[] Answer: \answerYes{}
    \item[] Justification: We attach a simplified version of our codebase as supplemental material. The accompanying \texttt{README.md} and \texttt{new\_dataset\_guide.md} provides step-by-step instructions for downloading the cross-matched Legacy--HSC sample from its public source (the Multimodal Universe dataset~\citep{themultimodaluniversecollaboration2024multimodaluniverseenablinglargescale}), preprocessing it, and training and evaluating the model. Both underlying galaxy imaging surveys (DESI Legacy Imaging Surveys and Hyper Suprime-Cam) are publicly accessible.

\item {\bf Experimental setting/details}
    \item[] Question: Does the paper specify all the training and test details (e.g., data splits, hyperparameters, how they were chosen, type of optimizer) necessary to understand the results?
    \item[] Answer: \answerYes{}
    \item[] Justification: Architecture, optimizer, hyperparameters, training schedule, and data splits are described in the methods section and the appendices on Model Architecture, Training Configuration, and the MLP probe used for downstream evaluation.

\item {\bf Experiment statistical significance}
    \item[] Question: Does the paper report error bars suitably and correctly defined or other appropriate information about the statistical significance of the experiments?
    \item[] Answer: \answerYes{}
    \item[] Justification: Statistical significance is reported where it is naturally available: $1\sigma$ posterior uncertainty bands estimated from 32 different samples per anchor are shown across spatial frequencies and lag distances in Fig.~\ref{fig:power_autocorr}, and pixel-wise $Z$-score distributions across all held-out galaxies are shown in Fig.~\ref{fig:generative_metrics}~(left). The downstream regression $R^2$ values in Table~\ref{tab:r2-scores} are point estimates from a single MLP probe trained per representation, with early stopping on a held-out validation set to control run-to-run variability. Given the size of the held-out evaluation set ($\sim$thousands of galaxies), the gaps between methods are substantially larger than the sampling-variability noise floor expected from a single $R^2$ estimate, and we do not expect multi-seed averaging to alter the qualitative ordering.

\item {\bf Experiments compute resources}
    \item[] Question: For each experiment, does the paper provide sufficient information on the computer resources (type of compute workers, memory, time of execution) needed to reproduce the experiments?
    \item[] Answer: \answerYes{}
    \item[] Justification: As reported in the Training Configuration appendix, the model was trained using distributed data parallel across 4 NVIDIA H100 GPUs with \textit{bfloat16} mixed precision, completing 75{,}000 training steps in approximately 8 hours of wall-clock time.
    
\item {\bf Code of ethics}
    \item[] Question: Does the research conducted in the paper conform, in every respect, with the NeurIPS Code of Ethics \url{https://neurips.cc/public/EthicsGuidelines}?
    \item[] Answer: \answerYes{}
    \item[] Justification: The research uses publicly available astronomical survey data and does not involve human subjects, sensitive data, or applications with foreseeable misuse potential.

\item {\bf Broader impacts}
    \item[] Question: Does the paper discuss both potential positive societal impacts and negative societal impacts of the work performed?
    \item[] Answer: \answerNA{}
    \item[] Justification: Our work targets representation learning for astrophysical galaxy imaging from public surveys. There is no foreseeable direct path to negative societal impacts from disentangling instrument systematics in scientific imaging.
    
\item {\bf Safeguards}
    \item[] Question: Does the paper describe safeguards that have been put in place for responsible release of data or models that have a high risk for misuse (e.g., pre-trained language models, image generators, or scraped datasets)?
    \item[] Answer: \answerNA{}
    \item[] Justification: The released code and any model checkpoints operate on astrophysical galaxy imaging from public surveys and pose no foreseeable risk of misuse.

\item {\bf Licenses for existing assets}
    \item[] Question: Are the creators or original owners of assets (e.g., code, data, models), used in the paper, properly credited and are the license and terms of use explicitly mentioned and properly respected?
    \item[] Answer: \answerYes{}
    \item[] Justification: We cite the original sources for all existing assets used: the DESI Legacy Imaging Surveys~\citep{Dey_2019}, Hyper Suprime-Cam~\citep{Aihara_2017}, the Multimodal Universe dataset~\citep{themultimodaluniversecollaboration2024multimodaluniverseenablinglargescale}, AION-1~\citep{parker2025aion1omnimodalfoundationmodel}, the ResNet architecture~\citep{he2015deepresiduallearningimage}, the \texttt{diffusers} \texttt{UNet2DConditionModel} implementation~\citep{von-platen-etal-2022-diffusers}, and the flow-matching framework~\citep{lipman2023flowmatchinggenerativemodeling}. All datasets used are publicly available under their respective survey data-release policies.

\item {\bf New assets}
    \item[] Question: Are new assets introduced in the paper well documented and is the documentation provided alongside the assets?
    \item[] Answer: \answerYes{}
    \item[] Justification: A simplified version of the codebase is submitted as supplemental material, with a \texttt{README} describing how to setup the dataset, train, and evaluate. The public code repository is linked in the first-page footnote.

\item {\bf Crowdsourcing and research with human subjects}
    \item[] Question: For crowdsourcing experiments and research with human subjects, does the paper include the full text of instructions given to participants and screenshots, if applicable, as well as details about compensation (if any)? 
    \item[] Answer: \answerNA{} 
    \item[] Justification: \answerNA{}

\item {\bf Institutional review board (IRB) approvals or equivalent for research with human subjects}
    \item[] Question: Does the paper describe potential risks incurred by study participants, whether such risks were disclosed to the subjects, and whether Institutional Review Board (IRB) approvals (or an equivalent approval/review based on the requirements of your country or institution) were obtained?
    \item[] Answer: \answerNA{} 
    \item[] Justification: \answerNA{}

\item {\bf Declaration of LLM usage}
    \item[] Question: Does the paper describe the usage of LLMs if it is an important, original, or non-standard component of the core methods in this research? Note that if the LLM is used only for writing, editing, or formatting purposes and does \emph{not} impact the core methodology, scientific rigor, or originality of the research, declaration is not required.
    \item[] Answer: \answerNA{}
    \item[] Justification: LLMs were not used as a component of the core methodology. Any use of LLMs was limited to writing, editing, or formatting and did not impact the scientific contributions of this work.

\end{enumerate}

\end{document}